\documentclass{LMCS}

\def\doi{9(3:4)2013}
\lmcsheading%
{\doi}
{1--28}
{}
{}
{Nov.~\phantom05, 2012}
{Aug.~14, 2013}
{}

\pdfoutput=1
\usepackage{enumerate}
\usepackage{hyperref}

\usepackage{amsmath}
\usepackage{amssymb}
\usepackage{xspace}
\usepackage{graphicx}

\newcommand{\ignore}[1]{}

\newcommand{\mc}[1]{\mathcal{#1}}
\newcommand{\df}{=_{\text{df}}}

\renewcommand{\epsilon}{\varepsilon}

\newcommand{\singleton}[1]{\{{#1}\}}
\newcommand{\setof}[2]{\{{#1}\,|\,{#2}\}}
\newcommand{\power}[1]{\mathcal{P}({#1})}
\newcommand{\npower}[1]{(\power{#1} \setminus \emptyset)}

\newcommand{\pair}[2]{({#1},{#2})}

\newcommand{\derives}[1]{\stackrel{#1}{\longrightarrow}}
\newcommand{\obsderives}[1]{\stackrel{#1}{\Longrightarrow}}

\newcommand{\mustderives}[1]{\stackrel{#1}{\longrightarrow}}
\newcommand{\mayderives}[1]{\stackrel{#1}{\dashrightarrow}}

\newcommand{\obsmayderives}[1]{\,\raisebox{1.0ex}{$\stackrel{#1}{\underset{\text{\normalsize $\dashrightarrow$}}{\raisebox{-1.0ex}[0ex][0ex]{$\dashrightarrow$}}}$}\,}

\newcommand{\obsmayderivesp}[2]{\,\raisebox{1.0ex}{$\stackrel{#1}{\underset{\text{\normalsize $\dashrightarrow$}}{\raisebox{-1.0ex}[0ex][0ex]{$\dashrightarrow$}}}$}_{#2}\,}

\newcommand{\fset}{F}
\newcommand{\falseff}{\textit{ff}}
\newcommand{\andop}{\wedge}
\newcommand{\orop}{\vee}
\newcommand{\cprod}{\&}
\newcommand{\parop}{|}
\newcommand{\pprod}{\otimes}

\newcommand{\iasim}{\sqsubseteq_{\textrm{IA}}}
\newcommand{\dmtssim}{\sqsubseteq_{\textrm{dMTS}}}
\newcommand{\dmtssiminv}{\sqsupseteq_{\textrm{dMTS}}}
\newcommand{\miasim}{\sqsubseteq_{\textrm{MIA}}}
\newcommand{\miaeq}{=_{\textrm{MIA}}}

\newcommand{\embed}[2]{[{#1}]_{\text{#2}}}
\newcommand{\univ}[1]{u_{#1}}

\begin{document}


\title[Modal Interface Automata]{Modal Interface Automata\rsuper*}


\author[G.~L{\"u}ttgen]{Gerald L{\"u}ttgen\rsuper a}
\address{{\lsuper a}Software Technologies Research Group,
         University of Bamberg,
         96045~Bamberg, Germany}
\email{gerald.luettgen@swt-bamberg.de}

\author[W.~Vogler]{Walter Vogler\rsuper b}
\address{{\lsuper b}Institute for Computer Science,
         University of Augsburg,
         86135~Augsburg, Germany}
\email{vogler@informatik.uni-augsburg.de}




\ACMCCS{[{\bf Theory of computation}]: Models of
  computation---Concurrency; Logic---Logic and Verification; Semantics
  and reasoning; [{\bf Software and its engineering}]: Context
  specific languages---Interface definition languages; Software
  system models---State systems}

\titlecomment{{\lsuper*}An extended abstract of this article appeared in 7th
  IFIP Intl.\ Conf.\ on \emph{Theoretical Computer Science}
  (TCS~2012), vol.~7604 of Lecture Notes in Computer Science,
  pp.~265--279, Springer, 2012.}


\begin{abstract}
  \noindent De~Alfaro and Henzinger's Interface Automata (IA) and
  Nyman et al.'s recent combination IOMTS of IA and Larsen's Modal
  Transition Systems (MTS) are established frameworks for specifying
  interfaces of system components.  However, neither IA nor IOMTS
  consider conjunction that is needed in practice when a component
  shall satisfy multiple interfaces, while Larsen's MTS-conjunction is
  not closed and Bene\v{s} et al.'s conjunction on disjunctive MTS
  does not treat internal transitions.  In addition, IOMTS-parallel
  composition exhibits a compositionality defect.

  This article defines conjunction (and also disjunction) on IA and
  disjunctive MTS and proves the operators to be `correct', i.e., the
  greatest lower bounds (least upper bounds) wrt.\ IA- and
  resp.\ MTS-refinement.  As its main contribution, a novel interface
  theory called Modal Interface Automata (MIA) is introduced: MIA is a
  rich subset of IOMTS featuring explicit output-must-transitions
  while input-transitions are always allowed implicitly, is equipped
  with compositional parallel, conjunction and disjunction operators,
  and allows a simpler embedding of IA than Nyman's.  Thus, it fixes
  the shortcomings of related work, without restricting designers to
  deterministic interfaces as Raclet et al.'s modal interface theory
  does.
\end{abstract}

\maketitle


\section{Introduction}
\label{sec:introduction}


Interfaces play an important role when designing complex software and
hardware systems so as to be able to check interoperability of system
components already at design stage.  Early interface theories deal
with types of data and operations only and have been successfully
deployed in compilers.  Over the past two decades, research has
focused on more advanced interface theories for \emph{sequential} and
object-oriented software systems, where interfaces also comprise
behavioural types.  Such types are often referred to as
\emph{contracts}~\cite{Mey92} and can express pre- and post-conditions
and invariants of methods and classes.  Much progress has been made on
the design of contract languages and on automated verification
techniques that can decide whether a system component meets its
contract (cf.~\cite{HatLeaLeinMuePar2012} for a survey).


More recently, \emph{behavioural interfaces} have also been proposed
and are being investigated for the use in \emph{concurrent} systems,
with prominent application examples being embedded systems
(e.g.,~\cite{MayGru2005}) and web services
(e.g.,~\cite{BeyChaHenSes2007, MerBjo2003}).  In this context,
behavioural interfaces are intended to capture protocol aspects of
component interaction.  One prominent example of such an interface
theory is de~Alfaro and Henzinger's \emph{Interface Automata}
(IA)~\cite{DeAHen2001, DeAHen2005}, which is based on labelled
transition systems (LTS) but distinguishes a component's input and
output actions.  The theory comes with an asymmetric parallel
composition operator, where a component may wait on inputs but never
on outputs.  Thus, a component's output must be consumed immediately,
or an error occurs.  In case no potential system environment may
restrict the system components' behaviour so that all errors are
avoided, the components are deemed to be incompatible.


Semantically, IA employs a refinement notion based on an alternating
simulation, such that a component satisfies an interface if (a)~it
implements all input behaviour prescribed by the interface and (b)~the
interface permits all output behaviour executed by the implementing
component.  Accordingly and surprisingly, an output in a specification
can always be ignored in an implementation.  In particular, a
component that consumes all inputs but never produces any output
satisfies any interface.  Since a specifier certainly wants to be able
to prescribe at least some outputs, Larsen, Nyman and Wasowski have
built their interface theory on Modal Transition Systems
(MTS)~\cite{Lar89} rather than LTS, which enables one to distinguish
between may- and must-transitions and thus to express mandatory
outputs.  The resulting \emph{IOMTS} interface
theory~\cite{LarNymWas2007}, into which IA can be embedded, is
equipped with an IA-style parallel composition and an MTS-style modal
refinement.  Unfortunately, IOMTS-modal refinement is not a
precongruence (i.e., not compositional) for parallel composition; a
related result in~\cite{LarNymWas2007} has already been shown
incorrect by Raclet et al.\ in~\cite{RacBadBenCaiLegPas2011}.


The present article starts from the observation that the above
interface theories are missing one important operator, namely
conjunction on interfaces.  Conjunction is needed in practice since
components are often designed to satisfy multiple interfaces
simultaneously, each of which specifies a particular aspect of
component interaction.  Indeed, conjunction is a key operator when
specifying and developing systems from different viewpoints as is
common in modern software engineering.
We thus start off by recalling the IA-setting and defining a
conjunction operator $\andop$ for IA; we prove that $\andop$ is indeed
conjunction, i.e., the greatest lower bound wrt.\ alternating
simulation (cf.\ Sec.~\ref{sec:ia}).  Essentially the same operator
has recently and independently been defined
in~\cite{CheChiJonKwi2012}, where it is shown that it gives the
greatest lower bound wrt.\ a \emph{trace-based} refinement relation.
As an aside, we also develop and investigate the dual disjunction
operator~$\orop$ for IA.  This is a natural operator for describing
alternatives in loose specifications, thus leaving implementation
decisions to implementors.

Similarly, we define conjunction and disjunction operators for a
slight extension of MTS (a subset of \emph{Disjunctive
  MTS}~\cite{LarXin90}, cf.\ Sec.~\ref{sec:dmts}), which paves us the
way for our main contribution outlined below.  Although Larsen has
already studied conjunction and disjunction for MTS, his operators do,
in contrast to ours, not preserve the MTS-property of syntactic
consistency, i.e., a conjunction or disjunction almost always has some
required transitions (must-transitions) that are not allowed (missing
may-transitions).  An additional difficulty when compared to the
IA-setting is that two MTS-interfaces may not have a common
implementation; indeed, inconsistencies may arise when composing MTSs
conjunctively.  We handle inconsistencies in a two-stage definition of
conjunction, adapting ideas from our prior work on conjunction in a
CSP-style process algebra~\cite{LueVog2010} that uses, however, a very
different parallel operator and refinement
preorder. In~\cite{BenCerKre2011}, a conjunction for Disjunctive MTS
(DMTS) is introduced in a two-stage style, too.  Our construction and
results for conjunction significantly extend the ones
of~\cite{BenCerKre2011} in that we also treat internal transitions
that, e.g., result from communication.

Note also that our setting employs event-based communication via
handshake and thus differs substantially from the one of shared-memory
communication studied by Abadi and Lamport in their paper on
conjoining specifications~\cite{AbaLam95}.  The same comment applies
to Doyen et al.~\cite{DoyHenJobPet2008}, who have studied a
conjunction operator for an interface theory involving shared-variable
communication.

Our article's main contribution is a novel interface theory, called
\emph{Modal Interface Automata} (MIA), which is essentially a rich
subset of IOMTS that still allows one to express
output-must-transitions.  In contrast to IOMTS, must-transitions can
also be disjunctive, and input-transitions are either required (i.e.,
must-transitions) or allowed implicitly.  MIA is equipped with an
MTS-style conjunction $\andop$, disjunction $\orop$ and an IOMTS-style
parallel composition operator, as well as with a slight adaptation of
IOMTS-refinement.  We show that (i)~MIA-refinement is a precongruence
for all three operators; (ii)~$\andop$ ($\orop$) is indeed conjunction
(disjunction) for this preorder; and (iii)~IA can be embedded into MIA
in a much cleaner, homomorphic fashion than into
IOMTS~\cite{LarNymWas2007} (cf.\ Sec.~\ref{sec:mia}).  Thereby, we
remedy the shortcomings of related work while, unlike the
language-based modal interface theory
of~\cite{RacBadBenCaiLegPas2011}, still permitting nondeterminism in
specifications.


\section{Conjunction and Disjunction for Interface Automata}
\label{sec:ia}

\emph{Interface Automata} (IA) were introduced by de~Alfaro and
Henzinger~\cite{DeAHen2001, DeAHen2005} as a \emph{reactive type}
theory that abstractly describes the communication behaviour of
software or hardware components in terms of their inputs and outputs.
IAs are labelled transition systems where visible actions are
partitioned into inputs and outputs.  The idea is that interfaces
interact with their environment according to the following rules.  An
interface cannot block an incoming input in any state but, if an input
arrives unexpectedly, it is treated as a catastrophic system failure.
This means that, if a state does not enable an input, this is a
requirement on the environment not to produce this input.  Vice versa,
an interface guarantees not to produce any unspecified outputs, which
are in turn inputs to the environment.

This intuition is reflected in the specific refinement relation of
\emph{alternating simulation} between IA and in the \emph{parallel
  composition} on IA, which have been defined in~\cite{DeAHen2005} and
are recalled in this section.  Most importantly, however, we introduce
and study a \emph{conjunction operator} on IA, which is needed in
practice to reason about components that are expected to satisfy
multiple interfaces.


\begin{defi}[Interface Automata~\cite{DeAHen2005}]
  An \emph{Interface Automaton} (IA) is a tuple $Q = (Q, I, O,
  \derives{})$, where
  \begin{enumerate}[(1)]
  \item $Q$~is a set of states,
  \item $I$ and~$O$ are disjoint input and output alphabets,
    resp., not containing the special, silent action~$\tau$,
  \item $\derives{} \,\subseteq Q \times (I \cup O \cup \{\tau\})
    \times Q$ is the \emph{transition relation}.
  \end{enumerate}
  The transition relation is required to be
  \emph{input-deterministic}, i.e., $a \in I$, $q \derives{a} q'$ and
  $q \derives{a} q''$ implies $q' = q''$.  In the remainder, we write
  $q \!\derives{a}$ if $q \derives{a} q'$ for some $q'$, as well as $q
  \,\not\!\derives{a}$ for its negation.
\label{def:ia}
\end{defi}
\noindent
In contrast to~\cite{DeAHen2005} we do not distinguish internal
actions and denote them all by~$\tau$, as is often done in process
algebras.
We let~$A$ stand for~$I \cup O$, let $a$ ($\alpha$) range over~$A$ ($A
\cup \{\tau\}$), and introduce the following weak transition
relations: $q \obsderives{\epsilon} q'$ if $q (\derives{\tau})^{\ast}
q'$, and $q \obsderives{o} q'$ for $o \in O$ if $\exists q''.\, q
\obsderives{\epsilon} q'' \derives{o} q'$; note that there are no
$\tau$-transitions after the $o$-transition.  Moreover, we define
$\hat{\alpha} = \epsilon$ if $\alpha = \tau$, and $\hat{\alpha} =
\alpha$ otherwise.


\begin{defi}[Alternating Simulation~\cite{DeAHen2005}]
    Let~$P$ and~$Q$ be IAs with common input and output alphabets.
    Relation $\mc{R} \subseteq P \times Q$ is an \emph{alternating
      simulation relation} if for all $\pair{p}{q} \in \mc{R}$:
  \begin{enumerate}[\bf(i):]
  \item $q \derives{a} q'$ and $a \in I$ implies $\exists p'.\, p
    \derives{a} p'$ and $\pair{p'}{q'} \in \mc{R}$,
  \item $p \derives{\alpha} p'$ and $\alpha \in O \cup \{\tau\}$
    implies $\exists q'.\, q \obsderives{\hat{\alpha}} q'$ and
    $\pair{p'}{q'} \in \mc{R}$.
  \end{enumerate}
  We write $p \iasim q$ and say that~$p$ \emph{IA-refines}~$q$ if
  there exists an alternating simulation relation~$\mc{R}$ such that
  $\pair{p}{q} \in \mc{R}$.
\label{def:iasim}
\end{defi}

\noindent
According to the basic idea of IA, if specification~$Q$ in state~$q$
allows some input~$a$ delivered by the environment, then the related
implementation state~$p$ of~$P$ must allow this input immediately in
order to avoid system failure.  Conversely, if~$P$ in state~$p$
produces output~$a$ to be consumed by the environment, this output
must be expected by the environment even if $q \obsderives{a}$; this
is because~$Q$ could have moved unobservedly from state~$q$ to
some~$q'$ that enables~$a$. Since inputs are not treated in
Def.~\ref{def:iasim} (ii), they are always allowed for~$p$.

It is easy to see that IA-refinement~$\iasim$ is a preorder on IA and
the largest alternating simulation relation.  Given input and output
alphabets~$I$ and~$O$, resp., the IA
\begin{equation*}
  \textit{BlackHole}_{I,O} \,\df\, (\{ \textit{blackhole} \}, I, O, \{
  (\textit{blackhole},a,\textit{blackhole}) \;|\; a \in I \})
\end{equation*}
IA-refines any other IA over~$I$ and~$O$.


\subsection{Conjunction on IA}
\label{subsec:iaconj}

Two IAs with common alphabets are always logically consistent in the
sense that they have a common implementation, e.g., the respective
blackhole IA as noted above.  This makes the definition of conjunction
on IA relatively straightforward. Here and similarly later, we index a
transition by the system's name to make clear from where it
originates, in case this is not obvious from the context.


\begin{defi}[Conjunction on IA]
  Let $P = (P, I, O, \derives{}_P)$ and $Q = (Q, I,$ $O,
  \derives{}_Q)$ be IAs with common input and output alphabets and
  disjoint state sets~$P$ and~$Q$.  The conjunction $P \andop Q$ is
  defined by $(\{ p \andop q \;|\; p \in P,\, q \in Q \} \cup P \cup
  Q, I, O, \derives{})$, where $\derives{}$ is the least set
  satisfying $\derives{}_P \subseteq \derives{}$, $\derives{}_Q
  \subseteq \derives{}$, and the following operational rules: \medskip

  \noindent
  \begin{tabular}{@{}l@{$\quad$}l@{$\!\quad$}l@{$\!\quad$}l@{}}
  {(I1)} &
  $p \andop q \derives{a} p'$ & if &
  $p \derives{a}_P p'$, $q \,\not\!\derives{a}_Q$ and $a \in I$
  \\
  {(I2)} &
  $p \andop q \derives{a} q'$ & if &
  $p \,\not\!\derives{a}_P$, $q \derives{a}_Q q'$ and $a \in I$
  \\
  {(I3)} &
  $p \andop q \derives{a} p' \andop q'$ & if &
  $p \derives{a}_P p'$, $q \derives{a}_Q q'$ and $a \in I$
  \\
  {(O)} &
  $p \andop q \derives{a} p' \andop q'$ & if &
  $p \derives{a}_P p'$, $q \derives{a}_Q q'$ and $a \in O$
  \\
  {(T1)} &
  $p \andop q \derives{\tau} p' \andop q$ & if &
  $p \derives{\tau}_P p'$
  \\
  {(T2)} &
  $p \andop q \derives{\tau} p \andop q'$ & if &
  $q \derives{\tau}_Q q'$
  \end{tabular}
\label{def:iaandop}
\end{defi}


\begin{figure}[tbp]
\phantom{.}\hfill%
\includegraphics[scale=0.52, clip=true]{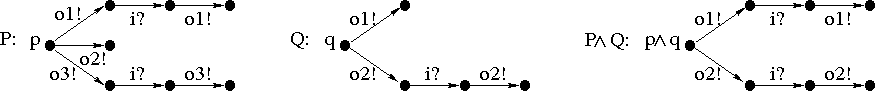}%
\hfill\phantom{.}
\caption{Example illustrating IA-conjunction.}
\label{fig:iaandopex}
\end{figure}

\noindent
Intuitively, conjunction is the synchronous product over actions
(cf.\ Rules~(I3), (O), (T1) and~(T2)).  Since inputs are always
implicitly present, this also explains Rules~(I1) and~(I2); for
example, in Rule~(I1), $q$ does not impose any restrictions on the
behaviour after input~$a$ and is therefore dropped from the target
state.
Moreover, the conjunction operator is commutative and associative.  As
an aside, note that the rules with digit~2 in their names are the
symmetric cases of the respective rules with digit~1; this convention
will hold true throughout this article.  Fig.~\ref{fig:iaandopex}
applies the rules above to an illustrating example; here and in the
following figures, we write~$a?$ for an input~$a$ and~$a!$ for an
output~$a$.

Essentially the same conjunction operator is defined by Chen et
al.\ in~\cite{CheChiJonKwi2012}, where a non-standard variant of IA is
studied that employs \emph{explicit} error states and uses a
trace-based semantics and refinement preorder (going back to
Dill~\cite{Dil89}).  The difference between their conjunction and
Def.~\ref{def:iaandop} is that error states are explicitly used in the
clauses that correspond to Rules~(I1) and~(I2) above, which renders
our definition arguably more elegant.  In~\cite{CheChiJonKwi2012}, an
analogue theorem to Thm.~\ref{thm:iaandisand} below is shown, but its
statement is different as it refers to a different refinement
preorder.  Also note that, deviating from the IA-literature, error
states are called inconsistent in~\cite{CheChiJonKwi2012}, but this is
not related to logic inconsistency as studied by us.

Our first result states that an implementation satisfies the
conjunction of interfaces exactly if it satisfies each of them.  This
is a desired property in system design where each interface describes
one aspect (or view) of the overall specification.


\begin{thm}[$\andop$ is And]
  Let $P, Q, R$ be IAs with states~$p$, $q$, $r$, resp.
  Then, $r \iasim p$ and $r \iasim q$ if and only if $r \iasim p
  \andop q$.
\label{thm:iaandisand}
\end{thm}

\proof{
  {``$\Longleftarrow$'':} It is sufficient to show that $\mc{R}
  \df \{ \pair{r}{p} \;|\; \exists q.\, r \iasim p \andop q \} \,\cup
  \iasim$ is an alternating simulation relation.  Let $\pair{r}{p} \in
  \mc{R}$ due to~$q$; the case $r \iasim p$ is obvious.  We check the
  conditions of Def.~\ref{def:iasim}:
  \begin{iteMize}{$\bullet$}
  \item Let $p \derives{a}_P p'$ with $a \in I$.
    \begin{iteMize}{$-$}
    \item\emph{$q \,\not\!\derives{a}_Q$:} Hence, $p \andop q
      \derives{a} p'$ by Rule~(I1) and, due to $r \iasim p \andop q$,
      there exists some~$r'$ with $r \derives{a}_R r'$ and $r' \iasim
      p'$.  Since $\pair{r'}{p'} \in \mc{R}$ we are done.

    \item\emph{$q \derives{a}_Q q'$:} Hence, $p \andop q \derives{a}
      p' \andop q'$ by Rule~(I3) and, due to $r \iasim p \andop q$,
      there exists some~$r'$ with $r \derives{a}_R r'$ and $r' \iasim
      p' \andop q'$.  Now, $\pair{p'}{q'} \in \mc{R}$.
    \end{iteMize}
  \item Let $r \derives{\alpha}_R r'$ with $\alpha \in O \cup \{\tau\}$.
    \begin{iteMize}{$-$}
    \item\emph{$\alpha \not= \tau$:} Thus, by Rule~(O) and possibly
      Rules~(T1), (T2), $p \andop q \obsderives{\alpha} p' \andop q'$
      with $r' \iasim p' \andop q'$.  We can project the transition
      sequence underlying $p \andop q \obsderives{\alpha} p' \andop
      q'$ to the $P$-component and get $p \obsderives{\alpha}_P p'$,
      and we are done since $\pair{r'}{p'} \in \mc{R}$.

    \item\emph{$\alpha = \tau$:} Hence, $p \andop q
      \obsderives{\epsilon} p' \andop q'$, possibly by Rules~(T1)
      and~(T2), with $r' \iasim p' \andop q'$.  Again, we can project
      to $p \obsderives{\epsilon}_P p'$ (where possibly $p' = p$) and
      also have $\pair{r'}{p'} \in \mc{R}$.
    \end{iteMize}
  \end{iteMize}
  \bigskip

  \noindent
  {``$\Longrightarrow$'':} We show that $\mc{R} \df \{ \pair{r}{p
    \andop q} \;|\; r \iasim p \text{ and } r \iasim q \} \,\cup \iasim$
  is an alternating simulation relation.  Let $\pair{r}{p} \in \mc{R}$;
  the case $r \iasim p$ is obvious, so we consider the following
  cases:
  \begin{enumerate}[(1)]
  \item$p \andop q \!\derives{a}$ with $a \in I$:
  \begin{enumerate}[\bf({I}1):]
  \item $p \andop q \derives{a} p'$ due to $p \derives{a}_P p'$
    and $q \,\not\!\derives{a}_Q$.  Then, $r \derives{a}_R r'$ for
    some~$r'$ with $r' \iasim p'$ due to $r \iasim p$, and we are done
    since $\pair{r'}{p'} \in \mc{R}$.

  \item Analogous to Case~(I1).

  \item $p \andop q \derives{a} p' \andop q'$ due to $p
    \derives{a}_P p'$ and $q \derives{a}_Q q'$.  Then, $r
    \derives{a}_R r'$ for some~$r'$ with $r' \iasim p'$ due to $r
    \iasim p$.  By input-determinism and $r \iasim q$, we also have
    $r' \iasim q'$ and are done since $\pair{r'}{p' \andop q'} \in
    \mc{R}$.
  \end{enumerate}

  \item$r \derives{\alpha}_R r'$ with $\alpha \in O \cup \{\tau\}$:
  \begin{iteMize}{$\bullet$}
  \item\emph{$\alpha \in O$:} Due to $r \iasim p$ and $r \iasim q$ we
    have~$p', q'$ such that $p \obsderives{\alpha}_P p'$, $q
    \obsderives{\alpha}_Q q'$, $r' \iasim p'$ and $r' \iasim q'$,
    i.e., $\pair{r'}{p' \andop q'} \in \mc{R}$.  We can interleave the
    $\tau$-transitions of the two transition sequences by Rules~(T1)
    and~(T2) and finally synchronize the two $\alpha$-transitions
    according to Rule~(O), and obtain $p \andop q \obsderives{\alpha}
    p' \andop q'$.

  \item\emph{$\alpha = \tau$:} Analogous, but without the synchronized
    transition.  \qed
  \end{iteMize}
  \end{enumerate}
}

\noindent
Technically, this result states that~$\andop$ gives the greatest
lower-bound wrt.\ $\iasim$ (up to equivalence), and its proof uses the
input-determinism property of IA.  The theorem also implies
compositional reasoning; from universal algebra one easily gets:

\begin{cor}
  For IAs $P, Q, R$ with states~$p$, $q$ and~$r$: $\,p \iasim q$
  $\;\Longrightarrow\;$ $p \andop r\iasim q \andop r$.
\label{cor:iaandopcomp}
\end{cor}

\proof{
  Assume $p \iasim q$.  Then, (always) $p \andop r \iasim p \andop r$
  $\Longleftrightarrow$ (by Thm.~\ref{thm:iaandisand}) $p \andop r
  \iasim p$ and $p \andop r \iasim r$ $\Longrightarrow$ (by assumption
  and transitivity) $p \andop r \iasim q$ and $p \andop r \iasim r$
  $\Longleftrightarrow$ (by Thm.~\ref{thm:iaandisand}) $p \andop r
  \iasim q \andop r$.  \qed
}



\subsection{Disjunction on IA}
\label{subsec:iadisj}

In analogy to conjunction we develop a disjunction operator on IA and
discuss its properties; in particular, this operator should give the
least upper bound.

\begin{defi}[Disjunction on IA]
  Let $P = (P, I, O, \derives{}_P)$ and $Q = (Q, I,$ $O,
  \derives{}_Q)$ be IAs with common input and output alphabets and
  disjoint state sets~$P$ and~$Q$.  The disjunction $P \orop Q$ is
  defined by $(\{ p \orop q \;|\; p \in P,\, q \in Q \} \cup P \cup Q,
  I, O, \derives{})$, where $\derives{}$ is the least set satisfying
  $\derives{}_P \subseteq \derives{}$, $\derives{}_Q \subseteq
  \derives{}$ and the following operational rules: \medskip

  \noindent
  \begin{tabular}{@{}l@{$\quad$}l@{$\!\quad$}l@{$\!\quad$}l@{}}
  {(I)} &
  $p \orop q \derives{a} p' \orop q'$ & if &
  $p \derives{a}_P p'$, $q \derives{a}_Q q'$ and $a \in I$
  \\
  {(OT1)} &
  $p \orop q \derives{\alpha} p'$ & if &
  $p \derives{\alpha}_P p'$ and $\alpha \in O \cup \{\tau\}$
  \\
  {(OT2)} &
  $p \orop q \derives{\alpha} q'$ & if &
  $q \derives{\alpha}_Q q'$ and $\alpha \in O \cup \{\tau\}$
  \end{tabular}
\label{def:iaorop}
\end{defi}

\noindent
Note that this definition preserves the input-determinism required of
IA.  The definition is roughly dual to the one of IA-conjunction,
i.e., we take the `intersection' of initial input behaviour and the
`union' of initial output behaviour.  Strictly speaking, this would
require the following additional rule for outputs~$o \in O$:\medskip

\noindent\begin{tabular}{@{}l@{$\quad$}l@{$\!\quad$}l@{$\!\quad$}l@{}}
  {(O3)} &
  $p \orop q \derives{o} p' \orop q'$ & if &
  $p \derives{o}_P p'$ and $q \derives{o}_Q q'$
\end{tabular}\medskip

\noindent However, the addition of this rule would in general result in
disjunctions $p \orop q$ that are larger than the least upper bound
of~$p$ and~$q$ wrt.~$\iasim$.  The following theorem shows that
our~$\orop$-operator properly characterizes the least upper bound:


\begin{thm}[$\orop$ is Or]
  Let $P, Q, R$ be IAs with states~$p$, $q$ and~$r$, resp.  Then, $p
  \orop q \iasim r$ if and only if $p \iasim r$ and $q \iasim r$.
\label{thm:iaorisor}
\end{thm}

\proof{
  \emph{``$\Longrightarrow$'':} We prove that $\mc{R} \df \{
  \pair{p}{r} \;|\; \exists q.\, p \orop q \iasim r \} \,\cup \iasim$
  is an alternating simulation relation.  We let $\pair{p}{r} \in
  \mc{R}$ due to~$q$ --~the case $p \iasim r$ is obvious~-- and check
  the conditions of Def.~\ref{def:iasim}:
  \begin{iteMize}{$\bullet$}
  \item Let $r \derives{a}_R r'$ with $a \in I$.  Hence, by $p \orop q
    \iasim r$ and the only applicable Rule~(I), $p \orop q \derives{a}
    p' \orop q'$ due to $p \derives{a}_P p'$ and $q \derives{a}_Q q'$
    with $p' \orop q' \iasim r'$.  Since $\pair{p'}{r'} \in \mc{R}$ we
    are done.

  \item Let $p \derives{\alpha}_P p'$ with $\alpha \in O \cup
    \{\tau\}$.  Hence, $p \orop q \derives{\alpha} p'$ by Rule~(OT1)
    and, due to $p \orop q \iasim r$, there exists some~$r'$ such that
    $r \obsderives{\hat{\alpha}} r'$ and $p' \iasim r'$.
  \end{iteMize}
  \bigskip

  \noindent
  {``$\Longleftarrow$'':} We show that $\mc{R} \df \{ \pair{p
    \orop q}{r} \;|\; p \iasim r \text{ and } q \iasim r \} \,\cup
  \iasim$ is an alternating simulation relation.  We let $\pair{p
    \orop q}{r} \in \mc{R}$ and consider the following cases:
  \begin{enumerate}[(1)]
  \item Let $r \derives{a}_R r'$ with $a \in I$.  By $p \iasim r$ and
    $q \iasim r$ we have~$p'$ and~$q'$ such that $p \derives{a}_P p'$,
    $q \derives{a}_Q q'$, $p' \iasim r'$ and $q' \iasim r'$.  Thus, we
    are done since $p \orop q \derives{a} p' \orop q'$ using Rule~(I)
    and since $\pair{p' \orop q'}{r'} \in \mc{R}$.

  \item $p \orop q \derives{\alpha} p'$ with $\alpha \in O \cup
    \{\tau\}$.  W.l.o.g., $p \derives{\alpha}_P p'$ due to Rule~(OT1).
    Then, $r \obsderives{\hat{\alpha}}_R r'$ for some~$r'$ satisfying
    $p' \iasim r'$, by $p \iasim r$.  \qed
  \end{enumerate}
}

\noindent
Compositionality of disjunction can now be derived dually to the proof
of Corollary~\ref{cor:iaandopcomp} but using Thm.~\ref{thm:iaorisor}
instead of Thm.~\ref{thm:iaandisand}:

\begin{cor}
  For IAs $P, Q, R$ with states~$p$, $q$ and~$r$: $\,p \iasim q$
  $\;\Longrightarrow\;$ $p \orop r\iasim q \orop r$.  \qed
\label{cor:iaoropcomp}
\end{cor}


\begin{figure}[tbp]
\phantom{.}\hfill%
\includegraphics[scale=0.52, clip=true]{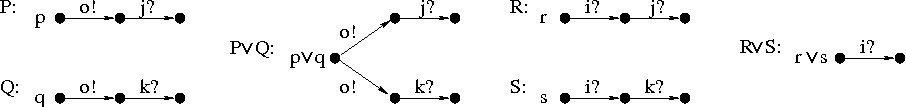}%
\hfill\phantom{.}
\caption{Example illustrating IA-disjunction's different treatment of
  inputs and outputs.}
\label{fig:iaoropex}
\end{figure}

\noindent
The two examples of Fig.~\ref{fig:iaoropex} round off our
investigation of IA disjunction by illustrating the operator's
different treatment of inputs and outputs.  Regarding $p \orop q$ on
the figure's left-hand side, the choice of which disjunct to implement
is taken with the first action~$o \in O$ if both disjuncts are
implemented; this meets the intuition of an inclusive-or.  In the
analogous situation of $r \orop s$ on the figure's right-hand side, a
branching on~$i \in I$ is not allowed due to input-determinism, and
the resulting IA is thus intuitively unsatisfactory.  The root cause
for this is that the IA-setting does not include sufficiently many
automata and, therefore, the least upper bound is `too large'.  The
shortcoming can be remedied by introducing disjunctive transitions, as
we will do below in the dMTS- and MIA-settings.  Then, we will have
more automata and, indeed, will get a smaller least upper
bound.\label{fromtena}


\subsection{Parallel Composition on IA}
\label{subsec:iaparop}

We recall the parallel composition operator~$\parop$ on IA
of~\cite{DeAHen2005}, which is defined in two stages: first a standard
product~$\pprod$ between two IAs is introduced, where common actions
are synchronized and hidden.  Then, error states are identified, and
all states are pruned from which reaching an error state is
unavoidable.


\begin{defi}[Parallel Product on IA~\cite{DeAHen2005}]
  IAs~$P_1$ and $P_2$ are called \emph{composable} if $A_1 \cap A_2 =
  (I_1 \cap O_2) \cup (O_1 \cap I_2)$, i.e., each common action is
  input of one IA and output of the other IA.  For such IAs we define
  the \emph{product} $P_1 \pprod P_2 = (P_1 \times P_2, I, O,
  \derives{})$, where $I = (I_1 \cup I_2) \setminus (O_1 \cup O_2)$
  and $O = (O_1 \cup O_2) \setminus (I_1 \cup I_2)$ and
  where~$\derives{}$ is given by the following operational
  rules: \medskip

  \noindent
  \begin{tabular}{@{}l@{$\quad$}l@{$\quad$}l@{$\quad$}l@{}}
  {(Par1)} &
  $\pair{p_1}{p_2} \derives{\alpha} \pair{p'_1}{p_2}$ & if &
  $p_1 \derives{\alpha} p'_1$ and $\alpha \notin A_2$
  \\
  {(Par2)} &
  $\pair{p_1}{p_2} \derives{\alpha} \pair{p_1}{p'_2}$ & if &
  $p_2 \derives{\alpha} p'_2$ and $\alpha \notin A_1$
  \\
  {(Par3)} &
  $\pair{p_1}{p_2} \derives{\tau} \pair{p'_1}{p'_2}$ & if &
  $p_1 \derives{a} p'_1$ and $p_2 \derives{a} p'_2$
  for some~$a$.
  \end{tabular}
\label{def:iaparprod}
\end{defi}

\noindent
Note that, in case of synchronization and according to Rule~(Par3),
one only gets internal $\tau$-transitions.

\begin{defi}[Parallel Composition on IA~\cite{DeAHen2005}]
  A state $\pair{p_1}{p_2}$ of a parallel product $P_1 \pprod P_2$ is
  an \emph{error state} if there is some $a \in A_1 \cap A_2$ such
  that (a)~$a \in O_1$, $p_1 \!\derives{a}$ and $p_2
  \,\not\!\derives{a}$, or (b)~$a \in O_2$, $p_2 \!\derives{a}$ and
  $p_1 \,\not\!\derives{a}$.

  A state of $P_1 \pprod P_2$ is \emph{incompatible} if it may reach
  an error state autonomously, i.e., only by output or internal
  actions that are, intuitively, locally controlled.  Formally, the
  set~$E \subseteq P_1 \times P_2$ of incompatible states is the least
  set such that $\pair{p_1}{p_2} \in E$ if (i)~$\pair{p_1}{p_2}$ is an
  error state or (ii)~$\pair{p_1}{p_2} \derives{\alpha}
  \pair{p'_1}{p'_2}$ for some $\alpha \in O \cup \{\tau\}$ and
  $\pair{p'_1}{p'_2} \in E$.

  The \emph{parallel composition}~$P_1 \parop P_2$ of~$P_1, P_2$ is
  obtained from $P_1 \pprod P_2$ by \emph{pruning}, i.e., removing all
  states in~$E$ and all transitions involving such states as source or
  target.  If $\pair{p_1}{p_2} \in P_1 \parop P_2$, we write~$p_1
  \parop p_2$ and call~$p_1$ and~$p_2$ \emph{compatible}.
\label{def:iaparop}
\end{defi}

\noindent
Parallel composition is well-defined since input-determinism is
preserved.


\begin{thm}[Compositionality of IA-Parallel Composition~\cite{DeAHen2005}]
  Let~$P_1$, $P_2$ and~$Q_1$ be IAs with $p_1 \in P_1$, $p_2 \in P_2$,
  $q_1 \in Q_1$ and $p_1 \iasim q_1$.  Assume that~$Q_1$ and~$P_2$ are
  composable; then, (a)~$P_1$ and~$P_2$ are composable and
  (b)~if~$q_1$ and~$p_2$ are compatible, then so are~$p_1$ and~$p_2$
  and $p_1 \parop p_2 \iasim q_1 \parop p_2$.  \qed
\label{thm:iaparopcomp}
\end{thm}

\noindent
This result relies on the fact that IAs are input-deterministic.
While the theorem is already stated in~\cite{DeAHen2005}, its proof is
only sketched therein.  Here, it is a simple corollary of
Thm.~\ref{thm:miaparopcomp} in Sec.~\ref{subsec:miaparop} and
Thms.~\ref{thm:iaembeddingmia} and~\ref{thm:miaembedding}(b) in
Sec.~\ref{subsec:embedding} below.


\begin{figure}[tbp]
\phantom{.}\hfill%
\includegraphics[scale=0.52, clip=true]{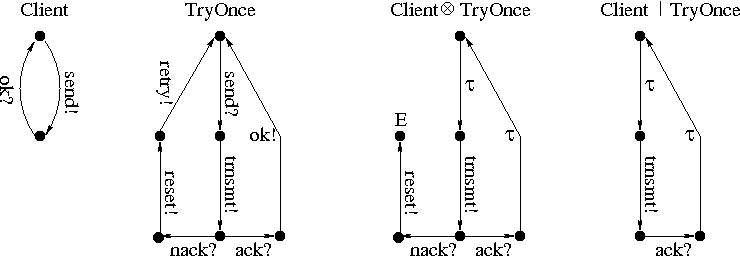}%
\hfill\phantom{.}
\caption{Example illustrating IA-parallel composition, where
  IA~\textit{TryOnce} has inputs $\{\textit{send, ack, nack}\}$ and
  outputs~$\{\textit{trnsmt, ok, reset, retry}\}$, while
  IA~\textit{Client} has inputs~$\{\textit{ok, retry}\}$ and
  outputs~$\{\textit{send}\}$.}
\label{fig:iaparopex}
\end{figure}

We conclude by presenting a small example of IA-parallel composition
in Fig.~\ref{fig:iaparopex}, which is adapted from~\cite{DeAHen2005}.
\textit{Client} does not accept its input \textit{retry}. Thus, if the
environment of $\textit{Client} \pprod \textit{TryOnce}$ would produce
\textit{nack}, the system would autonomously produce \textit{reset}
and run into a catastrophic error.  To avoid this, the environment of
$\textit{Client} \,\parop \textit{TryOnce}$ is required not to produce
\textit{nack}.
This view is called optimistic: there exists an environment in which
\textit{Client} and \textit{TryOnce} can cooperate without errors, and
$\textit{Client} \,\parop \textit{TryOnce}$ describes the necessary
requirements for such an environment.  In the pessimistic view as
advocated in~\cite{BauHenWir2011}, \textit{Client} and
\textit{TryOnce} are regarded as incompatible due to the potential
error.


\section{Conjunction and Disjunction for Modal Transition Systems}
\label{sec:dmts}


\emph{Modal Transition Systems} (MTS) were investigated by
Larsen~\cite{Lar89} as a specification framework based on labelled
transition systems but with two kinds of transitions: must-transitions
specify required behaviour, may-transitions specify allowed behaviour,
and absent transitions specify forbidden behaviour.  Any refinement of
an MTS-specification must preserve required and forbidden behaviour
and may turn allowed behaviour into required or forbidden behaviour.
Technically, this is achieved via an alternating-style simulation
relation, called \emph{modal refinement}, where any must-transition of
the specification must be simulated by an implementation, while any
may-transition of the implementation must be simulated by the
specification.


Our aim in this section is to extend MTS with conjunction and also
disjunction.  Larsen~\cite{Lar89} first defined conjunction and
disjunction on MTS (without $\tau$), but the resulting systems often
violate syntactic consistency (they are not really MTSs) and are hard
to understand.  This construction was subsequently generalized by
Larsen and Xinxin to Disjunctive MTS (DMTS)~\cite{LarXin90}, again
ignoring syntactic consistency.  This shortcoming was recently fixed
by Bene\v{s} et al.~\cite{BenCerKre2011} by exploiting the fact that
an $a$-must-transition in a DMTS may have several alternative target
states.  However, this work does still not consider a weak setting,
i.e., systems with~$\tau$.  Below, we will define conjunction and
disjunction on a syntactically consistent subclass of DMTS, called
\emph{dMTS}, but more generally in a weak setting as defined
in~\cite{DeAHen2005, LarNymWas2007}; this subclass is sufficient for
the purposes of the present article, and we leave the extension of our
results to DMTS for future work.  Since the treatment of
$\tau$-transitions is non-trivial and non-standard, we will motivate
and explain it in detail.


Note that this section will not consider parallel composition for
(d)MTS.  This is because we are working towards the MIA-setting that
will be introduced in the next section, which like IA and unlike
(d)MTS distinguishes between inputs and outputs.  (d)MTS parallel
composition can simply be defined in the style similar to
Def.~\ref{def:iaparprod}; in particular, it does not have error states
and thus fundamentally differs from conjunction as defined below.


\subsection{Disjunctive Modal Transition Systems}
\label{subsec:dmts}


We extend standard MTS only as far as needed for defining conjunction
and disjunction, by introducing disjunctive must-transitions that are
disjunctive wrt.\ exit states only (see Fig.~\ref{fig:dmtsandopex}).
The following extension also has no $\tau$-must-transitions since
these are not considered in the definition of the observational modal
refinement of~\cite{LarNymWas2007}.

\begin{defi}[disjunctive Modal Transition System]
  A \emph{disjunctive Modal Transition System} (dMTS) is a tuple $Q =
  (Q, A, \mustderives{}, \mayderives{})$, where
  \begin{enumerate}[(1)]
  \item $Q$~is a set of states,
  \item $A$~is an alphabet not containing the special, silent
        action~$\tau$,
  \item $\mustderives{} \,\subseteq Q \times A \times \npower{Q}$ is
        the \emph{must-transition} relation,
  \item $\mayderives{} \,\subseteq Q \times (A \cup \{\tau\}) \times Q$
        is the \emph{may-transition} relation.
  \end{enumerate}
  We require \emph{syntactic consistency}, i.e., $q \mustderives{a}
  Q'$ implies $\forall q' {\in} Q'.\, q \mayderives{a} q'$.
\label{def:dmts}
\end{defi}

\noindent
More generally, the must-transition relation in a standard
DMTS~\cite{LarXin90} may be a subset of $Q \times \npower{A \times
  Q}$.  For notational convenience, we write $q \mustderives{a} q'$
whenever $q \mustderives{a} \singleton{q'}$; all must-transitions in
standard MTS have this form.


Our refinement relation on dMTS abstracts from internal computation
steps in the same way as~\cite{LarNymWas2007}, i.e., by considering
the following \emph{weak may-transitions} for $\alpha \in A \cup
\{\tau\}$: $q \obsmayderives{\epsilon} q'$ if $q
\mayderives{\tau}^{\ast}\! q'$, and $q \obsmayderives{\alpha} q'$ if
$\exists q''.\, q \obsmayderives{\epsilon} q'' \mayderives{\alpha}
q'$.

\begin{defi}[Observational Modal Refinement, see~\cite{LarNymWas2007}]
  Let~$P, Q$ be dMTSs.  Relation $\mc{R} \subseteq P \times Q$ is an
  \emph{(observational) modal refinement relation} if for all
  $\pair{p}{q} \in \mc{R}$:
  \begin{enumerate}[\bf(i):]
  \item $q \mustderives{a} Q'$ implies $\exists P'.\,
    p \mustderives{a} P'$ and $\forall p' {\in} P'\,\exists q' {\in} Q'.\;
    \pair{p'}{q'} \in \mc{R}$,
  \item $p \mayderives{\alpha} p'$ implies $\exists q'.\, q
    \obsmayderives{\hat{\alpha}} q'$ and $\pair{p'}{q'} \in \mc{R}$.
  \end{enumerate}
  We write $p \dmtssim q$ and say that~$p$ \emph{dMTS-refines}~$q$ if
  there exists an observational modal refinement relation~$\mc{R}$ such
  that $\pair{p}{q} \in \mc{R}$.
\label{def:dmtssim}
\end{defi}

\noindent
Again, $\dmtssim$ is a preorder and the largest observational modal
refinement relation.  Except for disjunctiveness, dMTS-refinement is
exactly defined as for MTS in~\cite{LarNymWas2007}.  In the following
figures, any (disjunctive) must-transition drawn also represents
implicitly the respective may-transition(s), unless explicitly stated
otherwise.


\subsection{Conjunction on dMTS}
\label{subsec:dmtsconj}

Technically similar to parallel composition for IA, conjunction will
be defined in two stages. State pairs can be logically inconsistent
due to unsatisfiable must-transitions; in the second stage, we remove
such pairs incrementally.


\begin{defi}[Conjunctive Product on dMTS]
  Let $P = (P, A, \mustderives{}_P,$ $\mayderives{}_P)$ and $Q = (Q,
  A, \mustderives{}_Q, \mayderives{}_Q)$ be dMTSs with common
  alphabet.  The conjunctive product $P \cprod Q \df (P \times Q, A,
  \mustderives{}, \mayderives{})$ is defined by its operational
  transition rules as follows: \medskip

  \noindent
  \begin{tabular}{@{}l@{$\quad$}l@{$\!\quad$}l@{$\!\quad$}l@{}}
  {(Must1)} &
  $\pair{p}{q} \mustderives{a}
   \setof{\pair{p'}{q'}}{p' \in P',\, q \obsmayderivesp{a}{Q} q'}$ & if &
  $p \mustderives{a}_P P'$ and $q \obsmayderivesp{a}{Q}$
  \\
  {(Must2)} &
  $\pair{p}{q} \mustderives{a}
   \setof{\pair{p'}{q'}}{p \obsmayderivesp{a}{P} p',\, q' \in Q'}$ & if &
  $p \obsmayderivesp{a}{P}$ and $q \mustderives{a}_Q Q'$
  \\
  {(May1)} &
  $\pair{p}{q} \mayderives{\tau} \pair{p'}{q}$ & if &
  $p \obsmayderivesp{\tau}{P} p'$
  \\
  {(May2)} &
  $\pair{p}{q} \mayderives{\tau} \pair{p}{q'}$ & if &
  $q \obsmayderivesp{\tau}{Q} q'$
  \\
  {(May3)} &
  $\pair{p}{q} \mayderives{\alpha} \pair{p'}{q'}$ & if &
  $p \obsmayderivesp{\alpha}{P} p'$ and $q \obsmayderivesp{\alpha}{Q} q'$
  \end{tabular}
\label{def:dmtsconjprod}
\end{defi}

\noindent
It might be surprising that a single transition in the product might
stem from a transition sequence in one of the components (cf.\ the
first four items above) and that the components can also synchronize
on~$\tau$ (cf.\ Rule~(May3)).  The necessity of this is discussed
below; we only repeat here that conjunction is inherently different
from parallel composition where, for instance, there is no
synchronization on~$\tau$.

\begin{defi}[Conjunction on dMTS]
  Given a conjunctive product~$P \cprod Q$, the set $\fset \subseteq P
  \times Q$ of \emph{(logically) inconsistent states} is defined as
  the least set satisfying the following rules: \medskip

  \noindent
  \begin{tabular}{@{}l@{$\quad$}l@{$\quad$}l@{$\quad$}l@{}}
  {(F1)} &
  $p \!\mustderives{a}_P$, $q \not\!\!\!\obsmayderivesp{a}{Q}$ &
  implies &
  $\pair{p}{q} \in \fset$
  \\
  {(F2)} &
  $p \not\!\!\!\obsmayderivesp{a}{P}$, $q \!\mustderives{a}_Q$ &
  implies &
  $\pair{p}{q} \in \fset$
  \\
  {(F3)} &
  $\pair{p}{q} \mustderives{a} R'$ and $R' \subseteq \fset$ & implies &
  $\pair{p}{q} \in \fset$
  \end{tabular}
  \medskip

  \noindent
  The conjunction~$P \andop Q$ of dMTSs~$P, Q$ is obtained by deleting
  all states $\pair{p}{q} \in \fset$ from~$P \cprod Q$.  This also
  removes any may- or must-transition exiting a deleted state and any
  may-transition entering a deleted state; in addition, deleted states
  are removed from targets of disjunctive must-transitions. We write
  $p \andop q$ for the state~$\pair{p}{q}$ of~$P \andop Q$; these are
  the consistent states by construction, and $p \andop q$ is only
  defined for such a state.
\label{def:dmtsandop}
\end{defi}

\noindent
Regarding well-definedness, first observe that $P \cprod Q$ is a dMTS,
where syntactic consistency follows from Rule~(May3). Now, $P \andop
Q$ is a dMTS, too: if~$R'$ becomes empty for some $\pair{p}{q}
\mustderives{a} R'$, then also $\pair{p}{q}$ is deleted when
constructing $P \andop Q$ from $P \cprod Q$ according to~(F3).
Finally, our conjunction operator is also commutative and associative.


\begin{figure}[tbp]
\phantom{.}\hfill%
\includegraphics[scale=0.52, clip=true]{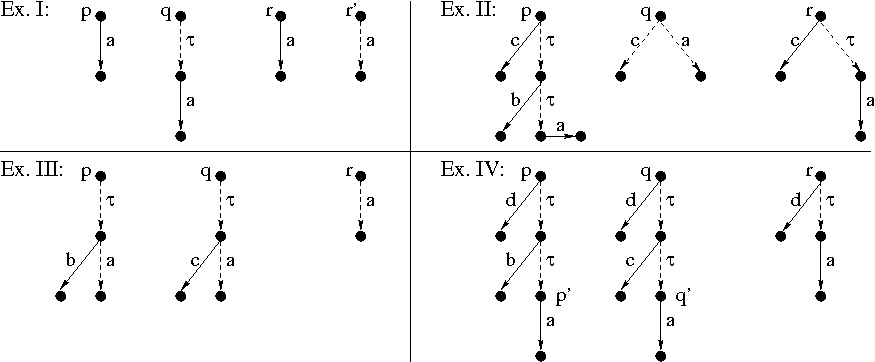}%
\hfill\phantom{.}
\caption{Examples motivating the rules of Def.~\ref{def:dmtsconjprod}.}
\label{fig:exdmtsconj}
\end{figure}

Before we formally state that operator~$\andop$ is indeed conjunction
on dMTS, we present several examples depicted in
Fig.~\ref{fig:exdmtsconj}, which motivate the rules of
Def.~\ref{def:dmtsconjprod}. In each case, $r$ is a common
implementation of $p$ and~$q$ (but not~$r'$ in Ex.~I), whence these
must be logically consistent.  Thus, Ex.~I explains Rule~(Must1).
If we only had~$\mayderives{\tau}$ in the precondition of Rule~(May1),
$p \andop q$ of~Ex.~II would just consist of a $c$-must- and an
$a$-may-transition; the only $\tau$-transition would lead to a state
in~$\fset$ due to $b$.  This would not allow the $\tau$-transition of
$r$, explaining Rule~(May1).
In Ex.~III and with only $\mayderives{\alpha}$ in the preconditions of
Rule~(May3), $p \andop q$ would just have three $\tau$-transitions to
inconsistent states (due to $b$, $c$, resp.).  This explains the weak
transitions for $\alpha \not= \tau$ in Rule~(May3).
According to Rules~(May1) and~(May2), $p \andop q$ in Ex.~IV has four
$\tau$-transitions to states in~$\fset$ (due to~$d$).  With
preconditions based on at least one~$\mayderives{\tau}$ instead
of~$\obsmayderives{\tau}$ in the $\tau$-case of Rule~(May3), there
would be three more $\tau$-transitions to states in~$\fset$ (due
to~$b$ or~$c$).  Thus, it is essential that Rule~(May3) also allows
the synchronization of two weak $\tau$-transitions, which in this case
gives $p \andop q \mayderives{\tau} p' \andop q'$.



\begin{figure}[tbp]
\phantom{.}\hfill%
\includegraphics[scale=0.52, clip=true]{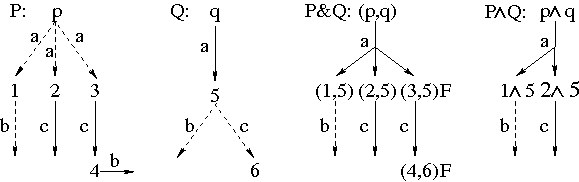}%
\hfill\phantom{.}
\caption{Example illustrating dMTS-conjunction.}
\label{fig:dmtsandopex}
\end{figure}

Fig.~\ref{fig:dmtsandopex} shows a small example illustrating the
treatment of disjunctive must-transitions in the presence of
inconsistency.  In~$P \cprod Q$, the $a$-must-transition of~$Q$
combines with the three $a$-transitions of~$P$ to a truly disjunctive
must-transition with a three-element target set.  The inconsistency of
state~$(4,6)$ due to~$b$ propagates back to state~$(3,5)$.  The
inconsistent states are then removed in~$P \andop Q$.


\begin{thm}[$\andop$ is And]
  Let $P, Q, R$ be dMTSs. Then, (i)~$(\exists r \in R.\, r \dmtssim p$
  and $r \dmtssim q)$ if and only if $p \andop q$ is defined.  In
  addition, in case $p \andop q$ is defined: (ii)~$r \dmtssim p$ and
  $r \dmtssim q \text{ if and only if } r \dmtssim p \andop q$.
\label{thm:dmtsandisand}
\end{thm}

\noindent
This key theorem states in Item~(ii) that conjunction behaves as it
should, i.e., $\andop$ on dMTSs is the greatest lower bound
wrt.~$\dmtssim$.  Item~(i) concerns the intuition that two
specifications~$p$ and~$q$ are logically inconsistent if they do not
have a common implementation; formally, $p \andop q$ is undefined in
this case.  Alternatively, we could have added an explicit
inconsistent element~$\falseff$ to our setting, so that $p \andop q =
\falseff$.  This element~$\falseff$ would be defined to be a
refinement of every~$p'$ and equivalent to any $\pair{p'}{q'} \in
\fset$ of some~$P \cprod Q$.  Additionally, $\falseff \andop p'$ and
$p' \andop \falseff$ would be defined as~$\falseff$, for any~$p'$.

The proof of the above theorem requires us to first introduce the
following concept for formally reasoning about inconsistent states:

\begin{defi}[dMTS-Witness]
  A \emph{dMTS-witness}~$W$ of~$P \cprod Q$ is a subset of~$P \times
  Q$ such that the following conditions hold for all $\pair{p}{q} \in
  W$:
  \medskip

  \noindent
  \begin{tabular}{@{}l@{$\quad$}l@{$\quad$}l@{$\quad$}l@{}}
  {(W1)} &
  $p \!\mustderives{a}_P$ & implies &
  $q \obsmayderivesp{a}{Q}$
  \\
  {(W2)} &
  $q \!\mustderives{a}_Q$ & implies &
  $p \obsmayderivesp{a}{P}$
  \\
  {(W3)} &
  $\pair{p}{q} \mustderives{a} R'$ & implies &
  $R' \cap W \not= \emptyset$
  \end{tabular}
\label{def:dmtswitness}
\end{defi}

\noindent
Conditions~(W1)--(W3) correspond to the negations of the premises of
Conditions~(F1)--(F3) in Def.~\ref{def:dmtsandop}.  This implies
Part~(i) of the following lemma, while Part~(ii) is essential for
proving Thm.~\ref{thm:dmtsandisand}(i):

\begin{lem}[Concrete dMTS-Witness]
  Let $P \cprod Q$ be a conjunctive product of dMTSs and $R$~be a
  dMTS.
  \begin{enumerate}[\bf(i):]
  \item For any dMTS-witness~$W$ of~$P \cprod Q$, we have
    $\fset \cap W = \emptyset$.
  \item The set $\{ \pair{p}{q} \in P \times Q \;|\; \exists
    r \in R.\, \text{$r \dmtssim p$}$ and $r \dmtssim q \}$ is a
    dMTS-witness of $P \cprod Q$.
  \end{enumerate}
\label{lem:dmtswitness}
\end{lem}

\proof{
  While the first statement of the lemma is quite obvious, we prove
  here that $W \df \{ \pair{p}{q} \in P \times Q \;|\; \exists r \in
  R.\, r \dmtssim p \text{ and } r \dmtssim q \}$ is a dMTS-witness of
  $P \cprod Q$ according to Def.~\ref{def:dmtswitness}:
  \begin{enumerate}[\hbox to8 pt{\hfill}]
  \item\noindent{\hskip-12 pt\bf(W1):}\ $p \mustderives{a}_P P'$ implies $r \mustderives{a}_R
    R'$ by $r \dmtssim p$.  Choose some $r' \in R'$.  Then, $r
    \mayderives{a}_R r'$ by syntactic consistency and $q
    \obsmayderivesp{a}{Q}$ by $r \dmtssim q$.

  \item\noindent{\hskip-12 pt\bf(W2):}\ Analogous to~(W1).

  \item\noindent{\hskip-12 pt\bf(W3):}\ Consider $\pair{p}{q} \in W$ due to~$r$, with
    $\pair{p}{q} \mustderives{a} S'$ due to $p \mustderives{a}_P P'$
    and $S' = \{ \pair{p'}{q'} \;|\; p' \in P',\, q
    \obsmayderivesp{a}{Q} q' \}$ according to Rule~(Must1).  By $r
    \dmtssim p$ we get some~$R' \subseteq R$ such that $r
    \mustderives{a}_R R'$ and $\forall r' {\in} R'\, \exists p' {\in}
    P'.\, r' \dmtssim p'$.  Choose $r' \in R'$; now, $r
    \mayderives{a}_R r'$ due to syntactic consistency, and $q
    \obsmayderivesp{a}{Q} q'$ with $r' \dmtssim q'$ by $r \dmtssim q$.
    Thus, we have $p' \in P'$ and $q'$~such that $\pair{p'}{q'} \in W
    \cap S'$ due to~$r'$. \qed
  \end{enumerate}
}

\noindent
We are now able to prove Thm.~\ref{thm:dmtsandisand}:

\proof{
  {(i)''$\Longrightarrow$'':} This follows from
  Lemma~\ref{lem:dmtswitness}. \medskip

  \noindent
  {(i), (ii)''$\Longleftarrow$'':} It suffices to show that
  $\mc{R} \df \{ \pair{r}{p} \;|\; \exists q.\, r \dmtssim p \andop q
  \}$ is an observational modal refinement relation.  Then, in
  particular, (i)''$\Longleftarrow$'' follows by choosing $r = p
  \andop q$.  We check the two conditions of Def.~\ref{def:dmtssim}:
  \begin{iteMize}{$\bullet$}
  \item Let $p \mustderives{a}_P P'$; then, $q \obsmayderivesp{a}{Q}$
    since, otherwise, $p \andop q$ would not be defined due to~(F1).
    Hence, by Rule~(Must1), $p \andop q \derives{a} \{ p' \andop q'
    \;|\; p' \in P',\, q \obsmayderivesp{a}{Q} q',$ \text{$p' \andop
      q'$} $\text{ defined} \}$.  By $r \dmtssim p \andop q$, we get
    $r \mustderives{a}_R R'$ such that $\forall r' {\in} R'$
    \text{$\exists p' \andop q'.$} $p' \in P'$, $q
    \obsmayderivesp{a}{Q} q'$ and $r' \dmtssim p' \andop q'$.  Hence,
    $\forall r' {\in} R'\, \exists p' {\in} P'.\, \pair{r'}{p'} \in
    \mc{R}$.

  \item $r \mayderives{\alpha}_R r'$ implies $\exists p' \andop q'.\,
    p \andop q \obsmayderives{\hat{\alpha}} p' \andop q'$ and $r'
    \dmtssim p' \andop q'$.  The contribution of~$p$ in this weak
    transition sequence gives $p \obsmayderivesp{\hat{\alpha}}{P} p'$,
    and we have $\pair{r'}{p'} \in \mc{R}$ due to~$q'$.
  \end{iteMize}
  \smallskip

  \noindent
  {(ii)''$\Longrightarrow$'':} Here, we show that $\mc{R} \df \{
  \pair{r}{p \andop q} \;|\; r \dmtssim p \text{ and } r \dmtssim q
  \}$ is an observational modal refinement relation.  By Part~(i), $p
  \andop q$ is defined and $\pair{r}{p \andop q}\in\mc{R}$ whenever $r
  \dmtssim p \text{ and } r \dmtssim q$.  We now verify the conditions
  of Def.~\ref{def:dmtssim}:
  \begin{iteMize}{$\bullet$}
  \item Let $p \andop q \mustderives{a} S'$, w.l.o.g.\ due to $p
    \mustderives{a}_P P'$ and $S' = \{ p' \andop q' \;|\; p' \in P',\,
    q \obsmayderivesp{a}{Q} q',$ \text{$p' \andop q' \text{ defined}
      \}$}.  Because of $r \dmtssim p$, we have $r \mustderives{a}_R
    R'$ so that \text{$\forall r' {\in} R'\, \exists p' {\in} P'.$}
    $r' \dmtssim p'$.  Consider some arbitrary $r' \in R'$ and the
    respective $p' \in P'$.  Then, $r \mayderives{a}_R r'$ by
    syntactic consistency and, due to $r \dmtssim q$, there exists
    some~$q'$ with $q \obsmayderivesp{a}{Q} q'$ and $r' \dmtssim q'$.
    Thus, $p' \andop q' \in S'$ and $\pair{r'}{p' \andop q'} \in
    \mc{R}$.

  \item Let $r \mayderives{\alpha}_R r'$ and consider $p
    \obsmayderivesp{\hat{\alpha}}{P} p'$ and $q
    \obsmayderivesp{\hat{\alpha}}{Q} q'$ satisfying $r' \dmtssim p'$
    and $r' \dmtssim q'$.  Thus, $\pair{r'}{p' \andop q'} \in \mc{R}$.
    Further, if $\alpha \not= \tau$, we have $p \andop q
    \mayderives{\alpha} p' \andop q'$ by Rule~(May3).  Otherwise,
    either $p \obsmayderivesp{\tau}{P} p'$ and $q
    \obsmayderivesp{\tau}{Q} q'$ and we are done by Rule~(May3) again,
    or w.l.o.g.\ $p \obsmayderivesp{\tau}{P} p'$ and $q=q'$ and we are
    done by Rule~(May1), or $p=p'$ and $q=q'$. \qed
  \end{iteMize}
}

\noindent
The following corollary of Thm.~\ref{thm:dmtsandisand} now easily
follows:

\begin{cor}
  dMTS-refinement is compositional wrt.\ conjunction, i.e., if $p
  \dmtssim q$ and $p \andop r$ is defined, then $q \andop r$ is
  defined and $p \andop r \dmtssim q \andop r$.
\label{cor:dmtsandopcomp}
\end{cor}

\proof{
  Assume $p \dmtssim q$ and $p \andop r$ is defined.  Then, (always)
  $p \andop r \dmtssim p \andop r$ $\Longleftrightarrow$ (by
  Thm.~\ref{thm:dmtsandisand}) $p \andop r \dmtssim p$ and $p \andop r
  \dmtssim r$ $\Longrightarrow$ (by assumption and transitivity) $p
  \andop r \dmtssim q$ and $p \andop r \dmtssim r$ $\Longrightarrow$
  (by Thm.~\ref{thm:dmtsandisand}(i)) $q \andop r$ is defined and (by
  Thm.~\ref{thm:dmtsandisand}(ii)) $p \andop r \dmtssim q \andop r$.
  \qed
}

\noindent
Thus, we have succeeded in our ambition to define a syntactically
consistent conjunction for MTS, for a weak MTS-variant with
disjunctive must-transitions.


\begin{figure}[tbp]
\phantom{.}\hfill%
\includegraphics[scale=0.54, clip=true]{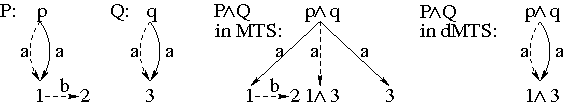}%
\hfill\phantom{.}
\caption{Example illustrating Larsen's MTS-conjunction;
  $\mayderives{a}$ drawn separately.}
\label{fig:larsen}
\end{figure}

Larsen~\cite{Lar89} also defines a conjunction operator on MTS, but
almost always the result violates syntactic consistency.  A simple
example is shown in Fig.~\ref{fig:larsen} where~$q$ refines~$p$ in
Larsen's setting as well as in our dMTS-setting; in this figure,
may-transitions are drawn explicitly, i.e, a must- is not necessarily
also a may-transition.  Since Larsen's $p \land q$ is not
syntactically consistent, this $p \land q$ and~$q$ are, contrary to
the first impression, equivalent.  In our dMTS-setting, $P \andop Q$
is isomorphic to~$Q$ which will also hold for our MIA-setting below
(with action~$b$ read as output and where $a$ could be either an input
or an output).


\begin{figure}[tbp]
\phantom{.}\hfill%
\includegraphics[scale=0.52, clip=true]{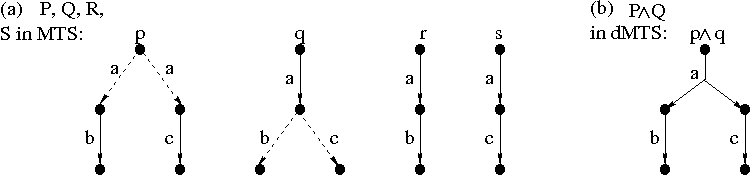}%
\hfill\phantom{.}
\caption{Example showing that conjunction cannot be defined on MTS.
  (A similar example is given in~\cite{BenCerKre2011} without proof.)}
\label{fig:mtsand}
\end{figure}

Indeed, conjunction cannot be defined on MTS in general, e.g., for
the~$P$ and~$Q$ in Fig.~\ref{fig:mtsand}(a).  The states~$p$ and~$q$
have~$r$ as well as~$s$ as common implementations; thus, $r$ and~$s$
must be implementations of~$p \andop q$.  An MTS~$P \andop Q$ would
need in state~$p \andop q$ (i)~an immediate $a$-must-transition (due
to~$q$) followed by (ii)~a must-$b$ and no~$c$ or a must-$c$ and
no~$b$ (due to~$p$).  In the first (second) case, $s$ ($r$) is not an
implementation of~$p \andop q$, which is a contradiction.  Using dMTS,
the conjunction~$P \andop Q$ is as shown in Fig.~\ref{fig:mtsand}(b).

The above shortcoming of MTS has been avoided by Larsen et
al.\ in~\cite{LarSteWei95} by limiting conjunction to so-called
\emph{independent} specifications that make inconsistencies obsolete;
this restriction also excludes the above example.
Recently, Bauer et al.~\cite{BauJuhLarLegSrb2012} have defined
conjunction for a version of MTS extended by partially ordered labels;
when refining an MTS, also the labels can be refined, and this has
various applications. However, the conjunction operator is only
defined under some restriction, which corresponds to requiring
determinism in the standard MTS-setting.
Another MTS-inspired theory including a conjunction operator has been
introduced by Raclet et al.~\cite{RacBadBenCaiLegPas2011}.  While
their approach yields the desired~$p \andop q$ as in our dMTS-setting,
it is language-based and thus deals with deterministic systems only.



\subsection{Disjunction on dMTS}
\label{subsec:dmtsdisj}

We will see in Sec.~\ref{subsec:iaembeddingdmts} that
input-transitions (output-trans\-itions) in IA correspond to
must-transitions (may-transitions) in dMTS.  In this light, the
following definition of disjunction corresponds closely to the one for
IA.  In particular, initial must-transitions are also combined, but
this time the choice between disjuncts is not delayed.

\begin{defi}[Disjunction on dMTS]
  Let $P = (P, A, \mustderives{}_P,$ $\mayderives{}_P)$ and $Q = (Q,
  A, \mustderives{}_Q,$ $\mayderives{}_Q)$ be dMTSs with common
  alphabet.  The disjunction $P \orop Q$ is defined as the tuple $(\{
  p \orop q \;|\; p \in P,\, q \in Q \} \cup P \cup Q, A,
  \mustderives{}, \mayderives{})$, where $\mustderives{}$ and
  $\mayderives{}$ are the least sets satisfying $\mustderives{}_P
  \subseteq \mustderives{}$, $\mayderives{}_P \subseteq
  \mayderives{}$, $\mustderives{}_Q \subseteq \mustderives{}$,
  $\mayderives{}_Q \subseteq \mayderives{}$ and the following
  operational rules: \medskip

  \noindent
  \begin{tabular}{@{}l@{$\quad$}l@{$\!\quad$}l@{$\!\quad$}l@{}}
  {(Must)} &
  $p \orop q \mustderives{a} P' \cup Q'$ & if &
  $p \mustderives{a}_P P'$, $q \mustderives{a}_Q Q'$
  \\
  {(May1)} &
  $p \orop q \mayderives{\alpha} p'$ & if &
  $p \mayderives{\alpha}_P p'$
  \\
  {(May2)} &
  $p \orop q \mayderives{\alpha} q'$ & if &
  $q \mayderives{\alpha}_Q q'$
  \end{tabular}
\label{def:dmtsorop}
\end{defi}


\noindent
This definition clearly yields well-defined dMTSs respecting syntactic
consistency.  It also gives us the desired least-upper-bound property:

\begin{thm}[$\orop$ is Or]
  Let $P$, $Q$, and~$R$ be dMTSs with states~$p$, $q$ and~$r$, resp.
  Then, $p \orop q \dmtssim r$ if and only if $p \dmtssim r$ and $q
  \dmtssim r$.
\label{thm:dmtsorisor}
\end{thm}

\proof{
  {``$\Longrightarrow$'':} We establish that $\mc{R} \df \{
  \pair{p}{r} \;|\; \exists q.\, p \orop q \dmtssim r \} \,\cup
  \dmtssim$ is a modal refinement relation.  To do so, we let
  $\pair{p}{r} \in \mc{R}$ due to~$q$ and check the conditions of
  Def.~\ref{def:dmtssim}:
  \begin{enumerate}[\bf(i):]
  \item Let $r \mustderives{a}_R R'$.  By $p \orop q \dmtssim r$ and
    the only applicable Rule~(Must), $p \orop q \mustderives{a} P'
    \cup Q'$ due to $p \mustderives{a}_P P'$ and $q \mustderives{a}_Q
    Q'$ such that $\forall p' {\in} P' \cup Q'\, \exists r' {\in}
    R'.\; p' \dmtssim r'$.  Hence, $\forall p' {\in} P'\, \exists r'
    {\in} R'.\; p' \dmtssim r'$ and, thus, $\pair{p'}{r'} \in \mc{R}$.

  \item Let $p \mayderives{\alpha}_P p'$.  Hence, $p \orop q
    \mayderives{\alpha} p'$ by Rule~(May1) and, due to $p \orop q
    \dmtssim r$, there exists some~$r'$ such that $r
    \obsmayderives{\hat{\alpha}}r'$ and $p' \dmtssim r'$.
  \end{enumerate}
  \bigskip

  \noindent
  {``$\Longleftarrow$'':} We prove that $\mc{R} \df \{ \pair{p
    \orop q}{r} \;|\; p \dmtssim r \text{ and } q \dmtssim r \} \,\cup
  \dmtssim$ is a modal refinement relation.  Let $\pair{p \orop q}{r}
  \in \mc{R}$ and consider the following cases:
  \begin{enumerate}[\bf(i):]
  \item Let $r \mustderives{a}_R R'$.  By $p \dmtssim r$ and $q
    \dmtssim r$ we have~$P'$, $Q'$ satisfying $p \mustderives{a}_P
    P'$, $q \mustderives{a}_Q Q'$ such that $\forall p' {\in} P'\,
    \exists r' {\in} R'.\; p' \dmtssim r'$ and $\forall q' {\in}
    Q'\, \exists r' {\in} R'.\; q' \dmtssim r'$.  Thus, $p \orop q
    \mustderives{a} P' \cup Q'$ using Rule~(Must) and we are done.

  \item $p \orop q \mayderives{\alpha} p'$.  W.l.o.g., this is
   due to Rule~(May1) and $p\mayderives{\alpha}_P p'$.  Then, $r
    \obsmayderives{\hat{\alpha}}_R r'$ for some~$r'$ satisfying $p'
    \dmtssim r'$, by $p \dmtssim r$.  \qed
  \end{enumerate}
}

\noindent
Analogously to the IA-setting we may obtain the following corollary to
the above theorem:

\begin{cor}
  dMTS-refinement is compositional wrt.\ disjunction.  \qed
\label{cor:dmtsoropcomp}
\end{cor}


\subsection{Embedding of IA into dMTS}
\label{subsec:iaembeddingdmts}


We can now adopt the embedding of IA into MTS
from~\cite{LarNymWas2007} to our setting:


\begin{defi}[IA-Embedding]
  Let~$P$ be an IA with $A = I \cup O$.  Then, the
  embedding~$\embed{P}{dMTS}$ of~$P$ into (d)MTS is defined as the
  (d)MTS $(P\cup\{\univ{P}\}, A, \mustderives{}, \mayderives{})$,
  where $\univ{P} \notin P$ and: \smallskip

  \noindent
  \begin{tabular}{@{$\qquad$}l@{$\quad$}l@{$\quad$}l@{}}
  $p \mayderives{\alpha} p'$ & if &
  $p \mustderives{\alpha}_P p'$ and $\alpha \in A \cup \{\tau\}$;
  \\
  $p \mustderives{a} p'$ & if &
  $p \mustderives{a}_P p'$ and $a \in I$;
  \\
  $p \mayderives{a} \univ{P}$ & if &
  $p \,\not\!\mustderives{a}_P$ and $a \in I$;
  \\
  $\univ{P} \mayderives{a} \univ{P}$ & if &
  $a \in A$.
  \end{tabular}
\label{def:iaembeddingdmts}
\end{defi}

\noindent
For the remainder of this section we simply write~$\embed{p}{}$ for $p
\in \embed{P}{dMTS}$.  Observe that $\embed{P}{dMTS}$ does not have
truly disjunctive transitions; hence, it is an MTS.
In~\cite{LarNymWas2007}, it is shown that this embedding respects
refinement, i.e., $p \iasim q$ if and only if $\embed{p}{} \dmtssim
\embed{q}{}$.  Since conjunction (disjunction) on IA and dMTS is the
greatest lower bound (least upper bound) wrt.~$\iasim$ and~$\dmtssim$
(up to equivalence), resp., we have by general order theory:

\begin{prop}[Conjunction/Disjunction and IA-Embedding]
  For all IAs~$P$ and~$Q$ with $p \in P$ and $q \in Q$:
  \begin{enumerate}[\bf(a):]
  \item
    $\embed{p \andop q}{} \,\dmtssim\, \embed{p}{} \andop \embed{q}{}$;
  \item
    $\embed{p \orop q}{} \,\dmtssiminv\, \embed{p}{} \orop \embed{q}{}$.
  \qed
  \end{enumerate}
\label{prop:iaembeddingdmts}
\end{prop}


\begin{figure}[tbp]
\phantom{.}\hfill%
\includegraphics[scale=0.52, clip=true]{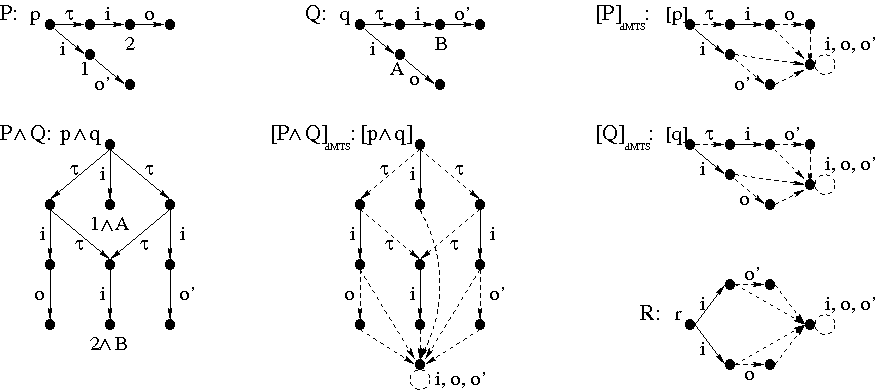}%
\hfill\phantom{.}
\caption{Example refuting the reverse refinement in
  Prop.~\ref{prop:iaembeddingdmts}(a).  All non-labelled transitions
  depict $i$-may-transitions.}
\label{fig:conjiaembeddingdmts}
\end{figure}


\begin{figure}[tbp]
\phantom{.}\hfill%
\includegraphics[scale=0.52, clip=true]{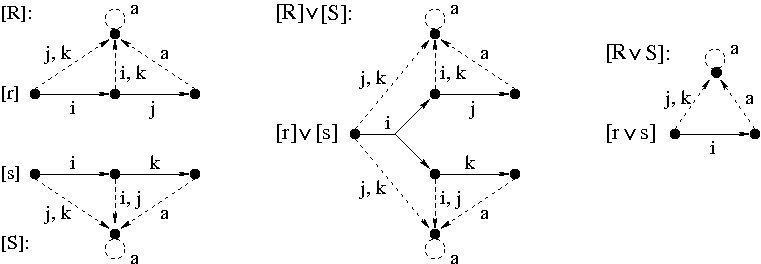}%
\hfill\phantom{.}
\caption{Example refuting the reverse refinement in
  Prop.~\ref{prop:iaembeddingdmts}(b) ($a \in A = \{i,j,k\}$).}
\label{fig:disjiaembeddingdmts}
\end{figure}

\noindent
The reverse refinements do not hold due to the additional dMTSs that
are not embeddings of IA.  To see this for conjunction, consider the
example in Fig.~\ref{fig:conjiaembeddingdmts}, where~$P$ and~$Q$ are
IAs.  State~$r$ in dMTS~$R$ is a common implementation of
state~$\embed{p}{}$ and state~$\embed{q}{}$, i.e., their conjunction
is sufficiently large to cover~$r$.  However, $r$~does not refine
$\embed{p \andop q}{}$ since the initial $i$-must-transition of the
latter cannot be matched by the former.  Hence, $\embed{p \andop q}{}$
and $\embed{p}{} \andop \embed{q}{}$ cannot be equivalent.  To see
this for disjunction, consider~$r$ and~$s$ in Fig.~\ref{fig:iaoropex}
on the right.  Fig.~\ref{fig:disjiaembeddingdmts} shows all relevant
dMTSs, and $\embed{r \orop s}{}$ does not refine~$\embed{r}{} \orop
\embed{s}{}$ since it does not have a must-transition after~$i$.


\section{Modal Interface Automata}
\label{sec:mia}

An essential point of Larsen, Nyman and Wasowski's
paper~\cite{LarNymWas2007} is to enrich IA with modalities to get a
flexible specification framework where inputs and outputs can be
prescribed, allowed or prohibited.  To do so, they consider IOMTS,
i.e., MTS where visible actions are partitioned into inputs and
outputs, and define parallel composition in IA-style.


\begin{figure}[tbp]
\phantom{.}\hfill%
\includegraphics[scale=0.52, clip=true]{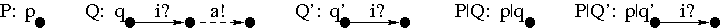}%
\hfill\phantom{.}
\caption{Example demonstrating the compositionality flaw of IOMTS.}
\label{fig:iomtsflaw}
\end{figure}

Our example of Fig.~\ref{fig:iomtsflaw} shows that their approach has
a serious flaw, namely observational modal refinement is not a
precongruence for the parallel composition of~\cite{LarNymWas2007}. In
this example, the IOMTS~$P$ has input alphabet~$\{a\}$ and empty
output alphabet, while~$Q$ and~$Q'$ have input alphabet~$\{i\}$ and
output alphabet~$\{a\}$.  Obviously, $q' \dmtssim q$.  When
composing~$P$ and~$Q$ in parallel, $p|q$ would reach an error state
after an $i$-must-transition in~\cite{LarNymWas2007} since the
potential output~$a$ of~$Q$ is not expected by~$P$.  In contrast,
$p|q'$ has an $i$-must- and $i$-may-transition not allowed by~$P|Q$,
so that $p|q' \not\dmtssim p|q$.  This counterexample also holds for
(strong) modal refinement as defined in~\cite{LarNymWas2007} and is
particularly severe since all systems are deterministic and all
must-transitions concern inputs only.  The problem is that~$p|q$
forbids input~$i$.

In~\cite{LarNymWas2007}, precongruence of parallel composition is not
mentioned.  Instead, a theorem relates the parallel composition of two
IOMTSs to a different composition on two refining implementations,
where an implementation in~\cite{LarNymWas2007} is an IOMTS in which
may- and must-transitions coincide.  This theorem is incorrect as is
pointed out in~\cite{RacBadBenCaiLegPas2011} and repaired in the
deterministic setting of that paper; the repair is again not a
precongruence result, but still compares the results of two different
composition operators.
However, a natural solution to the precongruence problem can be
adopted from the IA-framework~\cite{DeAHen2005} where inputs are
always allowed implicitly.  Consequently, if an input transition is
specified, it will always be a must.

In the remainder, we thus define and study a new specification
framework, called \emph{Modal Interface Automata} (MIA), that takes
the dMTS-setting for an alphabet consisting of input and output
actions, requires input-determinism, and demands that every
input-may-transition is also an input-must-transition.  The advantage
over IA is that outputs can be prescribed via output-must-transitions,
which precludes trivial implementations like \textit{BlackHole}
discussed in Sec.~\ref{sec:ia}.


\begin{defi}[Modal Interface Automaton]
  A \emph{Modal Interface Automaton} (MIA) is a tuple $Q = (Q, I, O,
  \mustderives{}, \mayderives{})$, where $(Q, I \cup O,
  \mustderives{}, \mayderives{})$ is a dMTS with disjoint
  alphabets~$I$ for inputs and~$O$ for outputs and where for all $i
  \in I$: (a)~$q \mustderives{i} Q'$ and $q \mustderives{i} Q''$ implies
  $Q' = Q''$; (b)~$q \mayderives{i} q'$ implies $\exists Q'.\, q
  \mustderives{i} Q'$ and $q' \in Q'$.
\label{def:mia}
\end{defi}

\noindent
In the conference version of this article, we have considered truly
disjunctive must-transitions only for outputs, so as to satisfy input
determinism; this suffices for developing MIA-conjunction.  However,
for disjunction we have seen that such transitions are also needed for
inputs.  The above definition of MIA therefore permits one disjunctive
must-transition for each input.  This allows some choice on performing
an input but, surprisingly, it is input-deterministic enough to
support compositionality for parallel composition
(cf.\ Thm.~\ref{thm:miaparopcomp}).


\begin{defi}[MIA-Refinement]
  Let~$P, Q$ be MIAs with common input and output alphabets.  Relation
  $\mc{R} \subseteq P \times Q$ is an \emph{(observational)
    MIA-refinement relation} if for all $\pair{p}{q} \in \mc{R}$:
  \begin{enumerate}[\bf(i):]
  \item $q \mustderives{a} Q'$ implies $\exists P'.\,
    p \mustderives{a} P'$ and $\forall p' {\in} P'\,\exists q' {\in} Q'.\;
    \pair{p'}{q'} \in \mc{R}$,
  \item $p \mayderives{\alpha} p'$ with $\alpha \in O \cup
    \singleton{\tau}$ implies $\exists q'.\, q
    \obsmayderives{\hat{\alpha}} q'$ and $\pair{p'}{q'} \in \mc{R}$.
  \end{enumerate}
  We write $p \miasim q$ and say that~$p$ \emph{MIA-refines}~$q$ if
  there exists an observational MIA-refinement relation~$\mc{R}$ such
  that $\pair{p}{q} \in \mc{R}$.  Moreover, we also write $p \miaeq q$
  in case $p \miasim q$ and $q \miasim p$ (which is an equivalence
  weaker than `bisimulation').
\label{def:miasim}
\end{defi}

\noindent
One can easily check that~$\miasim$ is a preorder and the largest
observational MIA-refinement relation.  Its definition coincides with
dMTS-refinement except that Cond.~(ii) is restricted to outputs and
the silent action~$\tau$.  Thus, inputs are always allowed implicitly
and, in effect, treated just like in IA-refinement.  Due to the
output-must-transitions in the MIA-setting, MIA-refinement can model,
e.g., STG-bisimilarity~\cite{VogWol2002} for systems without internal
actions; this is a kind of alternating simulation refinement used for
digital circuits.


\subsection{Conjunction on MIA}
\label{subsec:miaconj}

Similar to conjunction on dMTS, we define conjunction on MIA by first
constructing a conjunctive product and then eliminating all
inconsistent states.


\begin{defi}[Conjunctive Product on MIA]
  Let $P = (P, I, O, \mustderives{}_P,$ $\mayderives{}_P)$ and $Q =
  (Q, I, O, \mustderives{}_Q, \mayderives{}_Q)$ be MIAs with common
  input and output alphabets and disjoint state sets~$P$ and~$Q$.  The
  conjunctive product $P \cprod Q \df ((P \times Q) \cup P \cup Q, I,
  O, \mustderives{}, \mayderives{})$ inherits the transitions of~$P$
  and~$Q$ and has additional transitions as follows, where $i \in I$,
  $o \in O$ and $\alpha \in O \cup \{\tau\}$: \medskip

  \noindent
  \begin{tabular}{@{}l@{$\quad$}l@{$\!\quad$}l@{$\!\quad$}l@{}}
  {(OMust1)} &
  $\pair{p}{q} \mustderives{o}
   \setof{\pair{p'}{q'}}{p' \in P',\, q \obsmayderivesp{o}{Q} q'}$ & if &
  $p \mustderives{o}_P P'$ and $q \obsmayderivesp{o}{Q}$
  \\
  {(OMust2)} &
  $\pair{p}{q} \mustderives{o}
   \setof{\pair{p'}{q'}}{p \obsmayderivesp{o}{P} p',\, q' \in Q'}$ & if &
  $p \obsmayderivesp{o}{P}$ and $q \mustderives{o}_Q Q'$
  \\
  {(IMust1)} &
  $\pair{p}{q} \mustderives{i} P'$ & if &
  $p \mustderives{i}_P P'$ and $q \,\not\!\mustderives{i}_Q$
  \\
  {(IMust2)} &
  $\pair{p}{q} \mustderives{i} Q'$ & if &
  $p \,\not\!\mustderives{i}_P$ and $q \mustderives{i}_Q Q'$
  \\
  {(IMust3)} &
  $\pair{p}{q} \mustderives{i} P' \times Q'$ & if &
  $p \mustderives{i}_P P'$ and $q \mustderives{i}_Q Q'$
  \\
  {(May1)} &
  $\pair{p}{q} \mayderives{\tau} \pair{p'}{q}$ & if &
  $p \obsmayderivesp{\tau}{P} p'$
  \\
  {(May2)} &
  $\pair{p}{q} \mayderives{\tau} \pair{p}{q'}$ & if &
  $q \obsmayderivesp{\tau}{Q} q'$
  \\
  {(May3)} &
  $\pair{p}{q} \mayderives{\alpha} \pair{p'}{q'}$ & if &
  $p \obsmayderivesp{\alpha}{P} p'$ and
  $q \obsmayderivesp{\alpha}{Q} q'$
  \\
  {(IMay1)} &
  $\pair{p}{q} \mayderives{i} p'$ & if &
  $p \mayderives{i}_P p'$ and $q \,\not\!\mayderives{i}_Q$
  \\
  {(IMay2)} &
  $\pair{p}{q} \mayderives{i} q'$ & if &
  $p \,\not\!\mayderives{i}_P$ and $q \mayderives{i}_Q q'$
  \\
  {(IMay3)} &
  $\pair{p}{q} \mayderives{i} \pair{p'}{q'}$ & if &
  $p \mayderives{i}_P p'$ and $q \mayderives{i}_Q q'$
  \end{tabular}
\label{def:miaconjprod}
\end{defi}

\noindent
This product is defined analogously to IA-conjunction for inputs (plus
the corresponding `may' rules) and to the dMTS-product for outputs
and~$\tau$.  Thus, it combines the effects shown in
Fig.~\ref{fig:iaandopex} (where all outputs are treated as may) and
Fig.~\ref{fig:dmtsandopex} (where all actions are outputs).

\begin{defi}[Conjunction on MIA]
  Given a conjunctive product~$P \cprod Q$, the set $\fset \subseteq P
  \times Q$ of (logically) \emph{inconsistent states} is defined as
  the least set satisfying the following rules: \medskip

  \noindent
  \begin{tabular}{@{}l@{$\quad$}l@{$\quad$}l@{$\quad$}l@{}}
  {(F1)} &
  $p \!\mustderives{o}_P$, $q \not\!\!\!\obsmayderivesp{o}{Q}$, $o \in O$ &
  implies &
  $\pair{p}{q} \in \fset$
  \\
  {(F2)} &
  $p \not\!\!\!\obsmayderivesp{o}{P}$, $q \!\mustderives{o}_Q$, $o \in O$ &
  implies &
  $\pair{p}{q} \in \fset$
  \\
  {(F3)} &
  $\pair{p}{q} \mustderives{a} R'$ and $R' \subseteq \fset$ & implies &
  $\pair{p}{q} \in \fset$
  \end{tabular}
  \medskip

  \noindent
  The conjunction~$P \andop Q$ of MIAs~$P, Q$ with common input and
  output alphabets is obtained by deleting all states $\pair{p}{q} \in
  \fset$ from~$P \cprod Q$ as for dMTS in Def.~\ref{def:dmtsandop}.
  We write $p \andop q$ for state~$\pair{p}{q}$ of~$P \andop Q$; all
  such states are defined --~and consistent~-- by construction.
\label{def:miaandop}
\end{defi}

\noindent
The conjunction $P \andop Q$ is a MIA and is thus well-defined.  This
can be seen by a similar argument as we have used above in the context
of dMTS-conjunction, while input-determinism can be established by an
argument similar to that in the IA-setting.  Note that, in contrast to
the dMTS-situation, Rules~(F1) and~(F2) only apply to outputs.
Fig.~\ref{fig:dmtsandopex} is also an example for conjunction in the
MIA-setting if all actions are read as outputs.


To reason about inconsistency we use a notion of witness again.  This
may be defined analogously to the witness notion for dMTS but
replacing $a \in A$ in Def.~\ref{def:dmtswitness}(W1) and~(W2) by $a
\in O$.  We then obtain the analogous lemma to
Lemma~\ref{lem:dmtswitness}, which is needed in the proof of the
analogue theorem to Thm.~\ref{thm:dmtsandisand}:

\begin{defi}[MIA-Witness]
  A \emph{MIA-witness}~$W$ of~$P \cprod Q$ is a subset of $(P \times
  Q) \cup P \cup Q$ such that the following conditions hold for all
  $\pair{p}{q} \in W$:
  \medskip

  \noindent
  \begin{tabular}{@{}l@{$\quad$}l@{$\quad$}l@{$\quad$}l@{}}
  {(W1)} &
  $p \!\mustderives{o}_P$ with $o\in O$ & implies &
  $q \obsmayderivesp{o}{Q}$
  \\
  {(W2)} &
  $q \!\mustderives{o}_Q$  with $o\in O$ & implies &
  $p \obsmayderivesp{o}{P}$
  \\
  {(W3)} &
  $\pair{p}{q} \mustderives{a} R'$ & implies &
  $R' \cap W \not= \emptyset$
  \end{tabular}
\label{def:miawitness}
\end{defi}

\begin{lem}
  Let $P \cprod Q$ be a conjunctive product of MIAs.  Then, for any
  MIA-witness~$W$ of~$P \cprod Q$, we have (i)~$\fset \cap W =
  \emptyset$.  Moreover, (ii)~the set $W \df \{ \pair{p}{q} \in P
  \times Q \;|\; \exists\,\text{MIA}\,R$ and $r \in R.\, r \miasim p
  \text{ and } r \miasim q \} \cup P \cup Q$ is a MIA-witness of $P
  \cprod Q$.
\label{lem:miawitness}
\end{lem}


\proof{
  Since Part~(i) is again obvious, we directly proceed to proving
  Part~(ii), for which it suffices to consider the elements of $ \{
  \pair{p}{q} \in P \times Q \;|\; \exists r \in R.\, r \miasim p
  \text{ and } r \miasim q \}$; thus, let $\pair{p}{q} \in W$ due
  to MIA~$R$ and $r \in R$:
  \begin{enumerate}[\hbox to8 pt{\hfill}]
  \item\noindent{\hskip-12 pt\bf(W1):}\ $p \mustderives{o}_P P'$ implies $r \mustderives{o}_R
    R'$ by $r \miasim p$.  Choose some $r' \in R'$.  Then, $r
    \mayderives{o}_R r'$ by syntactic consistency, and $q
    \obsmayderivesp{o}{Q}$ by $r \miasim q$.

  \item\noindent{\hskip-12 pt\bf(W2):}\ Analogous to~(W1).

  \item\noindent{\hskip-12 pt\bf(W3):}\ Assume $\pair{p}{q} \!\mustderives{a}$.  According to
    the operational rules for conjunction, we distinguish the
    following cases:
    \begin{enumerate}[\hbox to8 pt{\hfill}]        
    \item\noindent{\hskip-12 pt\bf(OMust1):} Then, $\pair{p}{q}
      \mustderives{a} S'$ for $a \in O$, i.e., $p \mustderives{a}_P
      P'$ and $S' = \{ \pair{p'}{q'} \,|$ $p' \in P',\, q
      \obsmayderivesp{a}{Q} q' \}$.  By $r \miasim p$ we obtain
      some~$R' \subseteq R$ such that $r \mustderives{a}_R R'$ and
      $\forall r' {\in} R'$ $\exists p' {\in} P'.\, r' \miasim p'$.
      Choose $r' \in R'$ and the respective $p' \in P'$; now, $r
      \mayderives{a}_R r'$ due to syntactic consistency, and $q
      \obsmayderivesp{a}{Q} q'$ with $r' \miasim q'$ for some~$q'$ by
      $r \miasim q$.  Thus, we have $p' \in P'$ and $q'$~such that
      $\pair{p'}{q'} \in W \cap S'$ due to~$r'$.  Case~(OMust2) is
      analogous.

   \item\noindent{\hskip-12 pt\bf(IMust1):}\ Then, $\pair{p}{q} \mustderives{a} P'$ for $a \in
     I$, and we are done.  Case~(IMust2) is analogous.

   \item\noindent{\hskip-12 pt\bf(IMust3):}\ Then, $\pair{p}{q} \mustderives{a} P' \times Q'$
     for $a \in I$ due to $p \mustderives{a}_P P'$ and $q
     \mustderives{a}_Q Q'$.  By $r \miasim p$, $r \miasim q$ and
     input-determinism, we have some~$R'$ and $r' \in R'$ such that $r
     \mustderives{a}_R R'$, $\exists\, p' {\in} P'.\, r' \miasim p'$
     and $\exists\, q' {\in} Q'.\, r' \miasim q'$.  Thus,
     $\pair{p'}{q'} \in W$ due to~$r'$. \qed
    \end{enumerate}
  \end{enumerate}
}

\noindent
We can now state and prove the desired largest-lower-bound theorem,
from which compositionality of~$\miasim$ wrt.~$\andop$ follows in
analogy to the IA- and dMTS-settings:

\begin{thm}[$\andop$ is And]
  Let $P, Q$ be MIAs. We have \emph{(i)}~$(\exists\,\text{MIA}\,R$ and $r \in
  R.\, r \miasim p$ and $r \miasim q)$ if and only if $p \andop q$ is
  defined.  Further, in case $p \andop q$ is defined and for any
  MIA~$R$ and $r \in R$: \emph{(ii)}~$r \miasim p \text{ and } r \miasim q
  \text{ if and only if } r \miasim p \andop q$.
\label{thm:miaandisand}
\end{thm}

\proof{
  {(i)''$\Longrightarrow$'':} This follows directly from
  Lemma~\ref{lem:miawitness} above. \medskip

  \noindent
  {(ii)''$\Longleftarrow$'':} For a MIA~$R$ we show that $\mc{R}
  \df \{ \pair{r}{p} \in R \times P \;|\; \exists q \in Q.\; r \miasim
  p \andop q \} \,\cup \miasim$ is a MIA-refinement relation, by
  checking the two conditions of Def.~\ref{def:miasim} for some
  $\pair{r}{p} \in \mc{R}$ due to~$q$:
  \begin{iteMize}{$\bullet$}
  \item Let $p \mustderives{a}_P P'$ and consider the following cases
    depending on whether action~$a$ is an input or an output:
    \begin{iteMize}{$-$}
    \item{$a \in O$:}\ Then, $q \obsmayderivesp{a}{Q}$ since,
      otherwise, $p \andop q$ would not be defined due to~(F1).  Thus,
      by Rule~(OMust1), $p \andop q \derives{a} \{ p' \andop q' \;|\;
      p' \in P',\, q \obsmayderivesp{a}{Q} q',$ \text{$p' \andop q'$}
      $\text{defined} \}$.  By $r \miasim p \andop q$, we get some
      $R' \subseteq R$ such that $r \mustderives{a}_R R'$ and $\forall
      r' {\in} R'$ \text{$\exists p' \andop q'.$} $p' \in P'$, $q
      \obsmayderivesp{a}{Q} q'$ and $r' \miasim p' \andop q'$.  Hence,
      $\forall r' {\in} R'\, \exists p' {\in} P'.$ $\pair{r'}{p'} \in
      \mc{R}$.

    \item{$a \in I$:}\ This can lead to a transition of~$p \andop
      q$ in two ways:
      \begin{enumerate}[\hbox to8 pt{\hfill}]
      \item\noindent{\hskip-12 pt\bf (IMust1):}\ $q \,\not\!\mustderives{a}_Q$, whence $p \andop
        q \mustderives{a} P'$. By $r \miasim p \andop q$, there is
        some~$R'$ such that $r \mustderives{a}_R R'$ and $\forall r'
        {\in} R'\, \exists p' {\in} P'.\, r' \miasim p'$.

      \item\noindent{\hskip-12 pt\bf (IMust3):}\ $q \mustderives{a}_Q Q'$, whence $p \andop q
        \mustderives{a} (P' \times Q') \setminus \fset$.  By $r
        \miasim p \andop q$, there is some~$R'$ such that $r
        \mustderives{a}_R R'$ and $\forall r' {\in} R'\, \exists p'
        \andop q' \in P' \times Q'.\; r' \miasim p' \andop q'$ and,
        thus, $\pair{r'}{p'} \in \mc{R}$ due to~$q'$.
      \end{enumerate}
     \end{iteMize}

  \item $r \mayderives{\alpha}_R r'$ with $\alpha \in O \cup \{\tau\}$
    implies $\exists p' \andop q'.\, p \andop q
    \obsmayderives{\hat{\alpha}} p' \andop q'$ and $r' \miasim p'
    \andop q'$.  The contribution of~$p$ in this weak transition
    sequence gives $p \obsmayderivesp{\hat{\alpha}}{P} p'$, and we
    have $\pair{r'}{p'} \in \mc{R}$ due to~$q'$.
  \end{iteMize}
  \smallskip

  \noindent
  {(i)''$\Longleftarrow$'':}\ This follows
  from~(ii)''$\Longleftarrow$'' by choosing $R = P \andop Q$ and $r =
  p \andop q$.

  \medskip

  \noindent
  {(ii)''$\Longrightarrow$'':}\ Let~$R$ be a MIA~$R$.  We show
  that the relation $\mc{R} \df \{ \pair{r}{p \andop q} \;|\; r \in
  R,$ $r \miasim p \text{ and } r \miasim q \} \,\cup \miasim$ is a
  MIA-refinement relation.  Due to Part~(i), $p \andop q$ is defined
  whenever $r \miasim p$ and $r \miasim q$.  We now verify the
  conditions of Def.~\ref{def:miasim} for $\pair{r}{p \andop q} \in
  \mc{R}$:
  \begin{iteMize}{$\bullet$}
  \item Let $p \andop q \!\mustderives{a}$ and distinguish the
    following cases by our operational rules:
    \begin{iteMize}{$-$}
    \item$p \andop q \mustderives{a} S'$ with $a \in O$:  By Rule~(OMust1)
      this is
      w.l.o.g.\ due to $p \mustderives{a}_P P'$ and $S' =$ \text{$\{
        p' \andop q' \;|\; p' \in P',\, q \obsmayderivesp{a}{Q} q',$}
      \text{$p' \andop q' \text{ defined} \}$}.
      By $r \miasim p$, we have some $R' \subseteq R$ such that
      \text{$r \mustderives{a}_R R'$} and \text{$\forall r' {\in} R'$}
      $\exists p' {\in} P'.\, r' \miasim p'$.  Consider some arbitrary
      $r' \in R'$ and the respective $p' \in P'$.  Then, we have $r
      \mayderives{a}_R r'$ by syntactic consistency and, due to $r
      \miasim q$, some~$q'$ with $q \obsmayderivesp{a}{Q} q'$ and $r'
      \miasim q'$.  Thus, $p' \andop q' \in S'$ and $\pair{r'}{p'
        \andop q'} \in \mc{R}$.

    \item$p \andop q \mustderives{a} P'$ with $a \in I$: This is
      w.l.o.g.\ due to Rule~(IMust1): \text{$p \mustderives{a}_P P'$}
      and $q \,\not\!\mustderives{a}_Q$.  By $r \miasim p$, we have
      some~$R'$ such that $r \mustderives{a}_R R'$ and $\forall r'
      {\in} R'\, \exists p' {\in} P'.\, r' \miasim p'$, whence
      $\pair{r'}{p'} \in \mc{R}$.

    \item$p \andop q \mustderives{a} (P' \times Q') \setminus \fset$
      with $a \in I$: This is due to Rule~(IMust3), i.e., $p
      \mustderives{a}_P P'$ and $q \mustderives{a}_Q Q'$.  By $r
      \miasim p$ and $r \miasim q$, we get a unique~$r
      \mustderives{a}_R R'$ (by input-determinism) such that $\forall
      r' {\in} R'\, \exists p' {\in} P',\, q' {\in} Q'.\; r' \miasim
      p'$ and $r' \miasim q'$; thus, $\pair{r'}{p' \andop q'} \in
      \mc{R}$.
   \end{iteMize}

  \item Let $r \mayderives{\alpha}_R r'$ with $\alpha \in O \cup
    \{\tau\}$ and consider $p \obsmayderivesp{\hat{\alpha}}{P} p'$ and
    $q \obsmayderivesp{\hat{\alpha}}{Q} q'$ satisfying $r' \miasim p'$
    and $r' \miasim q'$.  Thus, $\pair{r'}{p' \andop q'} \in \mc{R}$.
    Further, if $\alpha \not= \tau$, we have $p \andop q
    \mayderives{\alpha} p' \andop q'$ by Rule~(May3).  Otherwise,
    either $p \obsmayderivesp{\tau}{P} p'$ and $q
    \obsmayderivesp{\tau}{Q} q'$ and we are done by Rule~(May3), or
    w.l.o.g.\ $p \obsmayderivesp{\tau}{P} p'$ and $q=q'$ and we are
    done by Rule~(May1), or $p=p'$ and $q=q'$. \qed
  \end{iteMize}
}

\noindent
In analogy to Corollary~\ref{cor:dmtsandopcomp} we obtain:

\begin{cor}
  MIA-refinement is compositional wrt.\ conjunction.  \qed
\label{cor:miaandopcomp}
\end{cor}


\subsection{Disjunction on MIA}
\label{subsec:miadisj}


The disjunction of two MIAs~$P$ and~$Q$ can be defined in the same way
as for dMTS, except for the special treatment of inputs in the
may-rules which guarantees that $P \orop Q$ is a MIA and, especially,
that Def.~\ref{def:mia}(b) is satisfied:

\begin{defi}[Disjunction on MIA]
  Let $P = (P, I, O, \mustderives{}_P,$ $\mayderives{}_P)$, $Q = (Q,
  I, O, \mustderives{}_Q,$ $\mayderives{}_Q)$ be MIAs with common
  input and output alphabets and disjoint state sets~$P$ and~$Q$.  The
  disjunction $P \orop Q$ is defined by $(\{ p \orop q \;|\; p \in
  P,\, q \in Q \} \cup P \cup Q, I, O, \mustderives{},
  \mayderives{})$, where $\mustderives{}$ and $\mayderives{}$ are the
  least sets satisfying $\mustderives{}_P \subseteq \mustderives{}$,
  $\mayderives{}_P \subseteq \mayderives{}$, $\mustderives{}_Q
  \subseteq \mustderives{}$, $\mayderives{}_Q \subseteq \mayderives{}$
  and the following operational rules: \medskip

  \noindent
  \begin{tabular}{@{}l@{$\quad$}l@{$\!\quad$}l@{$\!\quad$}l@{}}
  {(Must)} &
  $p \orop q \mustderives{a} P' \cup Q'$ & if &
  $p \mustderives{a}_P P'$ and $q \mustderives{a}_Q Q'$
  \\
  {(May1)} &
  $p \orop q \mayderives{\alpha} p'$ & if &
  $p \mayderives{\alpha}_P p'$ and,
  in case $\alpha \in I$, also $q \!\mayderives{\alpha}_Q$
  \\
  {(May2)} &
  $p \orop q \mayderives{\alpha} q'$ & if &
  $q \mayderives{\alpha}_Q q'$ and,
  in case $\alpha \in I$, also $p \!\mayderives{\alpha}_P$
  \end{tabular}
\label{def:miaorop}
\end{defi}


\noindent
It is easy to see that this definition is well-defined, i.e., the
resulting disjunctions are indeed MIAs, and we additionally have:

\begin{thm}[$\orop$ is Or]
  Let $P$, $Q$ and~$R$ be MIAs with states~$p$, $q$ and~$r$, resp.
  Then, $p \orop q \miasim r$ if and only if $p \miasim r$ and $q
  \miasim r$.  \qed
\label{thm:miaorisor}
\end{thm}

\noindent
The theorem's proof is as for dMTS (cf.\ Thm.~\ref{thm:dmtsorisor})
but, in the (ii)-cases, only $\alpha \in O \cup \{\tau\}$ has to be
considered.  Analogously to dMTS we obtain the following corollary to
Thm.~\ref{thm:miaorisor}:

\begin{cor}
  MIA-refinement is compositional wrt.\ disjunction.  \qed
\label{cor:miaoropcomp}
\end{cor}


\begin{figure}[tbp]
\phantom{.}\hfill%
\includegraphics[scale=0.52, clip=true]{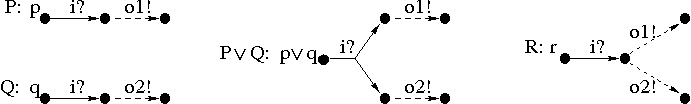}%
\hfill\phantom{.}
\caption{MIA-disjunction is more intuitive than IA-disjunction.}
\label{fig:miadisjintuitive}
\end{figure}

To conclude this section we argue that MIA-disjunction is more
intuitive than IA-disjunction. The example in
Fig.~\ref{fig:miadisjintuitive} shows MIAs~$P$, $Q$, $P \orop Q$ as
well as a MIA~$R$, where state~$r$ corresponds to the IA-disjunction
of states~$p$ and~$q$ when we understand~$P$ and~$Q$ as IAs.  As
expected (cf.\ p.~\pageref{fromtena}), $p \orop q$ is a refinement
of~$r$, but not vice versa.  MIA-disjunction can now be considered to
be more intuitive since the first transition in the disjunction
decides which disjunct has to be satisfied afterward, in contrast to
IA-disjunction.

\begin{figure}[tbp]
\phantom{.}\hfill%
\includegraphics[scale=0.52, clip=true]{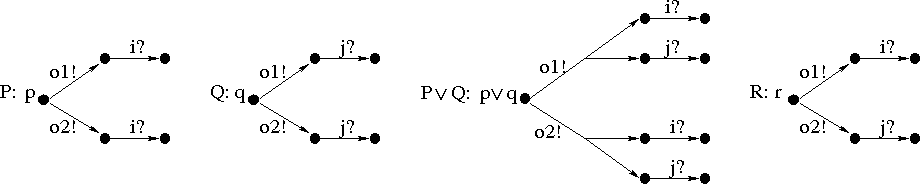}%
\hfill\phantom{.}
\caption{MIA-disjunction is an inclusive-or.}
\label{fig:miainclusiveor}
\end{figure}

Moreover, Fig.~\ref{fig:miainclusiveor} shows that MIA-disjunction is
an inclusive-or: an implementation of~$p \orop q$ can have an
$o1$-transition followed by~$i$ and another $o1$-transition followed
by~$j$; interestingly, $r \miasim p \orop q$ satisfies `half' of~$p$
and `half' of~$q$.  In general, for each action~\text{$a \in A$}
separately, a refinement of some disjunction has to satisfy at least
all initial $a$-must-transitions of one of its disjuncts.


\subsection{Parallel Composition on MIA}
\label{subsec:miaparop}

In analogy to the IA-setting~\cite{DeAHen2005} we provide a parallel
operator on MIA. Here, error states are identified, and all states are
removed from which reaching an error state is unavoidable in some
implementation, as is done for IOMTS in~\cite{LarNymWas2007}.


\begin{defi}[Parallel Product on MIA]
  MIAs~$P_1$ and~$P_2$ are \emph{composable} if $A_1 \cap A_2 = (I_1
  \cap O_2) \cup (O_1 \cap I_2)$, as in IA.  For such MIAs we define
  the \emph{product} \text{$P_1 \pprod P_2 = (P_1 \times P_2, I, O,$}
  $\mustderives{}, \mayderives{})$, where $I = (I_1 \cup I_2)
  \setminus (O_1 \cup O_2)$ and $O = (O_1 \cup O_2) \setminus (I_1
  \cup I_2)$ and where~$\mustderives{}$ and~$\mayderives{}$ are
  defined as follows: \medskip

  \noindent
  \begin{tabular}{@{}l@{$\quad$}l@{$\quad$}l@{$\quad$}l@{}}
  {(Must1)} &
  $\pair{p_1}{p_2} \mustderives{a} P'_1 \times \singleton{p_2}$ & if &
  $p_1 \mustderives{a} P'_1$ and $a \notin A_2$
  \\
  {(Must2)} &
  $\pair{p_1}{p_2} \mustderives{a} \singleton{p_1} \times P'_2$ & if &
  $p_2 \mustderives{a} P'_2$ and $a \notin A_1$
  \\
  {(May1)} &
  $\pair{p_1}{p_2} \mayderives{\alpha} \pair{p'_1}{p_2}$ & if &
  $p_1 \mayderives{\alpha} p'_1$ and $\alpha \notin A_2$
  \\
  {(May2)} &
  $\pair{p_1}{p_2} \mayderives{\alpha} \pair{p_1}{p'_2}$ & if &
  $p_2 \mayderives{\alpha} p'_2$ and $\alpha \notin A_1$
  \\
  {(May3)} &
  $\pair{p_1}{p_2} \mayderives{\tau} \pair{p'_1}{p'_2}$ & if &
  $p_1 \mayderives{a} p'_1$ and $p_2 \mayderives{a} p'_2$
  for some~$a$.
  \end{tabular}
\label{def:miaparprod}
\end{defi}

\noindent
Recall that there are no $\tau$-must-transitions since they are
irrelevant for refinement.

\begin{defi}[Parallel Composition on MIA]
  Given a parallel product $P_1 \pprod P_2$, a state $\pair{p_1}{p_2}$
  is an \emph{error state} if there is some $a \in A_1 \cap A_2$ such
  that (a)~$a \in O_1$, $p_1 \!\mayderives{a}$ and $p_2
  \,\not\!\mustderives{a}$, or (b)~$a \in O_2$, $p_2 \!\mayderives{a}$
  and $p_1 \,\not\!\mustderives{a}$.

  Again we define the set~$E \subseteq P_1 \times P_2$ of
  \emph{incompatible} states as the least set such that
  $\pair{p_1}{p_2} \in E$ if (i)~$\pair{p_1}{p_2}$ is an error state
  or (ii)~$\pair{p_1}{p_2} \mayderives{\alpha} \pair{p'_1}{p'_2}$ for
  some $\alpha \in O \cup \{\tau\}$ and $\pair{p'_1}{p'_2} \in E$.

  The \emph{parallel composition}~$P_1 \parop P_2$ of~$P_1$ and~$P_2$
  is now obtained from $P_1 \pprod P_2$ by \emph{pruning}, namely
  removing all states in~$E$ and every transition that involves such
  states as its source, its target or one of its targets; all
  may-transitions underlying a removed must-transition are deleted,
  too.  If $\pair{p_1}{p_2} \in P_1 \parop P_2$, we write~$p_1 \parop
  p_2$ and call~$p_1$ and~$p_2$ \emph{compatible}.
\label{def:miaparop}
\end{defi}


\noindent
Parallel products and parallel compositions are well-defined MIAs.
Syntactic consistency is preserved, as is input-determinism since
input-transitions are directly inherited from one of the
\emph{composable} systems.  In particular, Cond.~(b) in
Def.~\ref{def:mia} holds due to the additional clause regarding the
deletion of may-transitions.  In addition, targets of disjunctive
must-transitions are never empty since all must-transitions that
remain after pruning are taken from the product without modification.


As an example why pruning is needed, consider Fig.~\ref{fig:iaparopex}
again and read the $\tau$-transitions as may-transitions and all other
transitions as must-transitions.  Further observe that pruning is
different from removing inconsistent states in conjunction.  For truly
disjunctive transitions $\pair{p_1}{p_2} \mustderives{a} P'$ of the
product $P_1 \pprod P_2$, the state $\pair{p_1}{p_2}$ is removed
already if $P' \cap E \not= \emptyset$, i.e., there exists some
$\pair{p'_1}{p'_2} \in P' \cap E$, and not only if $P' \subseteq E$.
This is clear for $a \in O$ since $\pair{p_1}{p_2} \mayderives{a}
\pair{p'_1}{p'_2}$ by syntactic consistency and, therefore,
$\pair{p_1}{p_2}$ is deleted itself by Cond.~(ii) above.  Note that
Cond.~(ii) corresponds directly to the IA-case since
output-transitions there correspond to may-transitions here (see
Sec.~\ref{subsec:iaembeddingdmts}).  For $a \in I$, reaching the error
state can only be prevented if the environment does not provide~$a$;
intuitively, this is because~$P'$ has w.l.o.g.\ the form~$P'_1 \times
\{p_2\}$ in the product of~$P_1$ and~$P_2$ (i.e., $p'_2 = p_2$).  The
implementor of~$P_1$ might choose to implement $p_1 \derives{a} p'_1$
such that --~when~$P_1$'s implementation is composed with~$P_2$'s~--
the error state is reached.  To express the requirement on the
environment not to exhibit~$a$, must-transition $\pair{p_1}{p_2}
\mustderives{a} P'$ and all underlying may-transitions have to be
deleted.


\begin{thm}[Compositionality of MIA-Parallel Composition]
  Let~$P_1$, $P_2$ and~$Q_1$ be MIAs with $p_1 \in P_1$, $p_2 \in
  P_2$, $q_1 \in Q_1$ and $p_1 \miasim q_1$.  Assume that~$Q_1$
  and~$P_2$ are composable; then:
  \begin{enumerate}[\bf(a):]
  \item $P_1$ and~$P_2$ are composable.
  \item If~$q_1$ and~$p_2$ are compatible, then so are~$p_1$,
    $p_2$ and $p_1 \parop p_2 \miasim q_1 \parop p_2$.
  \end{enumerate}
\label{thm:miaparopcomp}
\end{thm}


\proof{
  Part~(a) follows immediately since MIA~$Q_1$ has the same input and
  output alphabets as MIA~$P_1$, due to~$p_1 \miasim q_1$.  Regarding
  Part~(b), the first claim is implied by the following auxiliary
  result:
  \begin{quote}
    Let~$E_P$ be the $E$-set of~$P_1 \pprod P_2$ and~$E_Q$ be the one
    of~$Q_1 \pprod P_2$.  Then, $\pair{p_1}{p_2} \in E_P$ and $p_1
    \miasim q_1$ together imply $\pair{q_1}{p_2} \in E_Q$.
  \end{quote}
  The proof of this result is by induction on the length of a path
  from~$\pair{p_1}{p_2}$ to an error state of~$P_1 \pprod P_2$:
  \begin{enumerate}[\hbox to8 pt{\hfill}]  
  \item\noindent{\hskip-12 pt\bf (Base):}\ Let $\pair{p_1}{p_2}$ be an error state.
  \begin{iteMize}{$\bullet$}
  \item Let $p_1 \!\mayderives{a}_{P_1}$ with $a \in O_1 \cap I_2$ and
    $p_2 \,\not\!\mustderives{a}_{P_2}$.  Then, for some~$q'_1$, we
    have $q_1 \obsmayderives{\epsilon}_{Q_1} q'_1
    \!\mayderives{a}_{Q_1}$ by $p_1 \miasim q_1$; therefore,
    $\pair{q_1}{p_2} \obsmayderives{\epsilon} \pair{q'_1}{p_2} \in
    E_Q$ and $\pair{q_1}{p_2} \in E_Q$, too.

  \item Let $p_2 \!\mayderives{a}_{P_2}$ with $a \in O_2 \cap I_1$ and
    $p_1 \,\not\!\mustderives{a}_{P_1}$.  If $q_1
    \!\mustderives{a}_{Q_1}$, we have a contradiction to~$p_1 \miasim
    q_1$; otherwise, $\pair{q_1}{p_2}$ is an error state.
  \end{iteMize}

  \item\noindent{\hskip-12 pt\bf (Step):}\ For a shortest path from~$\pair{p_1}{p_2}$ to an error
    state, consider the first transition $\pair{p_1}{p_2}
    \mayderives{\alpha} \pair{p'_1}{p'_2} \in E_P$ with $\alpha \in O
    \cup \{\tau\}$.  The transition is due to Rule~(May1), (May2)
    or~(May3).  In all cases we show $p'_1 \miasim q'_1$, which
    implies $\pair{q'_1}{p'_2} \in E_Q$ by induction hypothesis.
  \begin{enumerate}[\hbox to8 pt{\hfill}]  
  \item\noindent{\hskip-12 pt\bf (May1):}\ $p_1 \mayderives{\alpha}_{P_1} p'_1$, $p_2 = p'_2$,
    $\alpha \notin A_2$, and $\alpha \in O_1 \cup \{\tau\}$ by $\alpha
    \in O \cup \{\tau\}$.  Hence, there is some~$q'_1$ such that $q_1
    \obsmayderives{\hat{\alpha}}_{Q_1} q'_1$ and $p'_1 \miasim q'_1$,
    due to $p_1 \miasim q_1$, and $\pair{q_1}{p_2}
    \obsmayderives{\hat{\alpha}} \pair{q'_1}{p_2}$ by applications of
    Rule~(May1).  By induction hypothesis, $\pair{q'_1}{p_2} \in E_Q$
    and, thus, $\pair{q_1}{p_2} \in E_Q$.

  \item\noindent{\hskip-12 pt\bf (May2):}\ $p_1 = p'_1$, $p_2 \mayderives{\alpha}_{P_2} p'_2$ and
    $\alpha \notin A_1$.  Now, since~$P_1$ and~$Q_1$ have the same
    alphabets by~$p_1 \miasim q_1$, we can apply Rule~(May2) again and
    obtain $\pair{q_1}{p_2} \mayderives{\alpha} \pair{q_1}{p'_2}$, so
    that~$\pair{q_1}{p'_2} \in E_Q$ by induction hypothesis.  Hence,
    $\pair{q_1}{p_2} \in E_Q$, too.

  \item\noindent{\hskip-12 pt\bf (May3):}\ $\alpha = \tau$.
    \begin{iteMize}{$\bullet$}
    \item $p_1 \mayderives{a}_{P_1} p'_1$ with $a \in O_1$, and $p_2
      \mayderives{a}_{P_2} p'_2$ with $a \in I_2$.  By $p_1 \miasim
      q_1$, we have $q_1 \obsmayderives{\epsilon}_{Q_1} q''_1
      \mayderives{a}_{Q_1} q'_1$ for some~$q'_1, q''_1$ with $p'_1
      \miasim q'_1$.  Hence, $\pair{q_1}{p_2} \obsmayderives{\epsilon}
      \pair{q''_1}{p_2} \mayderives{\tau} \pair{q'_1}{p'_2}$ via
      Rules~(May1) and~(May3).  By induction hypothesis,
      $\pair{q'_1}{p'_2} \in E_Q$ and, thus, $\pair{q_1}{p_2} \in
      E_Q$, too.

    \item $p_1 \mayderives{a}_{P_1} p'_1$ with $a \in I_1$, and $p_2
      \mayderives{a}_{P_2} p'_2$ with $a \in O_2$.  If $q_1
      \,\not\!\mayderives{a}_{Q_1}$, then $q_1
      \,\not\!\mustderives{a}_{Q_1}$ by syntactic consistency and
      $\pair{q_1}{p_2}$ is thus an error state.  If $q_1
      \!\mayderives{a}_{Q_1}$, then there exist unique $p_1
      \mustderives{a}_{P_1} P'$ and $q_1 \mustderives{a}_{Q_1} Q'$.
      We have \text{$p'_1 \in P'$} by Def.~\ref{def:mia}(b) and
      $\exists q'_1 {\in} Q'.\, p'_1 \miasim q'_1$ since $p_1 \miasim
      q_1$.  Hence, $q_1 \mayderives{a}_{Q_1} q'_1$ by syntactic
      consistency and $\pair{q_1}{p_2} \mayderives{\tau}
      \pair{q'_1}{p'_2}$ due to Rule~(May3).  By induction hypothesis,
      $\pair{q'_1}{p'_2} \in E_Q$ and, therefore, $\pair{q_1}{p_2} \in
      E_Q$.
    \end{iteMize}
  \end{enumerate}
  \end{enumerate}

  \noindent This completes the proof of the auxiliary result.  We can now prove
  that
  \begin{equation*}
  \mc{R} \,\df\, \{ \pair{p_1 \parop p_2}{q_1 \parop p_2} \;|\;
  p_1 \miasim q_1, \text{ $p_1, p_2$ as well as $q_1, p_2$ compatible} \}
  \end{equation*}
  is a MIA-refinement relation, for which we let $\pair{p_1 \parop
    p_2}{q_1 \parop p_2} \in \mc{R}$ and check the conditions of
  Def.~\ref{def:miasim}:
  \begin{enumerate}[\bf(i):]
  \item Let $q_1 \parop p_2 \mustderives{a} Q'$ with $Q' \cap E_Q
    = \emptyset$ due to either Rule~(Must1) or~(Must2).
    \begin{enumerate}[\hbox to8 pt{\hfill}]                                         \item\noindent{\hskip-12 pt\bf (Must1):}\ $q_1 \mustderives{a}_{Q_1} Q'_1$ and $Q' = Q'_1
      \times \{p_2\}$.  Then, by $p_1 \miasim q_1$, there is some
      $P'_1 \subseteq P_1$ such that $p_1 \mustderives{a}_{P_1} P'_1$
      and $\forall p'_1 {\in} P'_1\, \exists q'_1 {\in} Q'_1.\, p'_1
      \miasim q'_1$.  Now, $\pair{p_1}{p_2} \mustderives{a} P'_1
      \times \{p_2\}$ by Rule~(Must1) and since $a \notin A_2$.  For
      $p'_1 \in P'_1$ we have a suitable $q'_1 \in Q_1'$, and
      $\pair{p'_1}{p_2} \notin E_P$ since $\pair{q'_1}{p_2} \notin
      E_Q$ and due to the auxiliary result above.  Thus, for the
      arbitrary~$p'_1 \parop p_2$, we also have $\pair{p_1' \parop
        p_2}{q_1' \parop p_2} \in \mc{R}$.

    \item\noindent{\hskip-12 pt\bf (Must2):}\ $p_2 \mustderives{a}_{P_2} P'_2$ and $Q' = \{q_1\}
      \times P'_2$.  Then, $\pair{p_1}{p_2} \derives{a} P' = \{p_1\}
      \times P'_2$ by Rule~(Must2) and as~$P_1, Q_1$ have the same
      alphabets by~$p_1 \miasim q_1$.  For $\pair{p_1}{p'_2} \in P'$,
      we get $\pair{p_1}{p'_2} \notin E_P$ since $\pair{q_1}{p'_2}
      \notin E_Q$ and due to the auxiliary result above.  Thus, $p_1
      \parop p_2 \mustderives{a} P'$ and, for $p_1 \parop p'_2 \in
      P'$, we have $q_1 \parop p'_2 \in Q'$ with $\pair{p_1 \parop
        p'_2}{q_1 \parop p'_2} \in \mc{R}$.
    \end{enumerate}
  \item Let $p_1 \parop p_2 \mayderives{\alpha} p'_1 \parop p'_2
    \notin E_P$ with $\alpha \in O \cup \{\tau\}$.  The transition
    arises from one of the Rules~(May1), (May2) or~(May3):
    \begin{enumerate}[\hbox to8 pt{\hfill}]  
    \item\noindent{\hskip-12 pt\bf (May1):}\ $p'_2 = p_2$ and $p_1 \mayderives{\alpha}_{P_1}
      p'_1$.  By $p_1 \miasim q_1$, we have $q_1
      \obsmayderives{\hat{\alpha}}_{Q_1} q'_1$ for some~$q'_1$ such
      that $p'_1 \miasim q'_1$.  Hence, $\pair{q_1}{p_2}
      \obsmayderives{\hat{\alpha}} \pair{q'_1}{p_2}$ by repeated
      application of Rule~(May1) and since $\omega \notin A_2$.  If
      any state on this transition sequence were in~$E_Q$, then also
      $\pair{q_1}{p_2} \in E_Q$ which contradicts $\pair{p_1 \parop
        p_2}{q_1 \parop p_2} \in \mc{R}$.  Thus, $q_1 \parop p_2
      \obsmayderives{\hat{\alpha}} q'_1 \parop p_2$ with $\pair{p'_1
        \parop p_2}{q'_1 \parop p_2} \in \mc{R}$.

    \item\noindent{\hskip-12 pt\bf (May2):}\ $p'_1 = p_1$ and $p_2 \mayderives{\alpha}_{P_2}
      p'_2$.  Then, $\pair{q_1}{p_2} \mayderives{\alpha}
      \pair{q_1}{p'_2}$ by Rule~(May2) and since~$P_1$ and~$Q_1$ have
      the same alphabets due to~$p_1 \miasim q_1$.  If the latter
      state~$\pair{q_1}{p'_2}$ were in~$E_Q$, then also the former
      state~$\pair{q_1}{p_2}$.  Therefore, we have $q_1 \parop p_2
      \mayderives{\alpha} q_1 \parop p'_2$ and, moreover, $\pair{p_1
        \parop p'_2}{q_1 \parop p'_2} \in \mc{R}$.

    \item\noindent{\hskip-12 pt\bf (May3):}\ $\alpha = \tau$, $p_1 \mayderives{a}_{P_1} p'_1$ and
      $p_2 \mayderives{a}_{P_2} p'_2$ for some~$a$.
      \begin{iteMize}{$\bullet$}
      \item{$a \in O_1 \cap I_2$:} Then, $q_1
        \obsmayderives{\epsilon}_{Q_1} q''_1 \mayderives{a}_{Q_1}
        q'_1$ for~$q'_1, q''_1$ with $p'_1 \miasim q'_1$, due to $p_1
        \miasim q_1$.  Now, $\pair{q_1}{p_2} \obsmayderives{\epsilon}
        \pair{q''_1}{p_2} \mayderives{\tau} \pair{q'_1}{p'_2}$ by
        Rules (May1), (May3).  As in Case~(May1) above, $q_1 \parop
        p_2 \obsmayderives{\epsilon} q'_1 \parop p'_2$ and $\pair{p'_1
          \parop p'_2}{q'_1 \parop p'_2} \in \mc{R}$.

      \item{$a \in I_1 \cap O_2$:} If $q_1
        \,\not\!\mayderives{a}_{Q_1}$, then $\pair{q_1}{p_2}$ would be
        an error state, which is a contradiction.  Therefore, $q_1
        \!\mayderives{a}_{Q_1}$ and, by Def.~\ref{def:mia}(b), there
        exist unique $p_1 \mustderives{a}_{P_1} P'$ and $q_1
        \mustderives{a}_{Q_1} Q'$ by input-determinism.  We have $p'_1
        \in P'$ and $\exists q'_1 {\in} Q'.\, p'_1 \miasim q'_1$ since
        $p_1 \miasim q_1$.  Thus, $\pair{q_1}{p_2} \mayderives{\tau}
        \pair{q'_1}{p'_2}$ by Rule~(May3) and syntactic consistency,
        and $\pair{q'_1}{p'_2} \notin E_Q$ by the same reasoning as
        above.  Hence, $q_1 \parop p_2 \mayderives{\tau} q'_1 \parop
        p'_2$ with $\pair{p'_1 \parop p'_2}{q'_1 \parop p'_2} \in
        \mc{R}$.  \qed
      \end{iteMize}
    \end{enumerate}
  \end{enumerate}
}

\begin{figure}[tbp]
\phantom{.}\hfill%
\includegraphics[scale=0.52, clip=true]{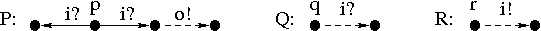}%
\hfill\phantom{.}
\caption{Example illustrating the need of input-determinism for MIA.}
\label{fig:miainputdet}
\end{figure}

\noindent
This precongruence property of MIA-refinement would not hold if we
would do away with input-determinism in MIA.  To see this, consider
the example of Fig.~\ref{fig:miainputdet} for which $p \miasim q$;
however, $p \parop r \miasim q \parop r$ does not hold since~$q$
and~$r$ are compatible while~$p$ and~$r$ are not.  An analogue
reasoning applies to IA, although we do not know of a reference in the
IA literature where this has been observed.


\subsection{Embedding of IA into MIA}
\label{subsec:embedding}

To conclude, we provide an embedding of IA into MIA in the line
of~\cite{LarNymWas2007}:


\begin{defi}[IA-Embedding]
  Let~$P$ be an IA.  The embedding~$\embed{P}{MIA}$ of~$P$ into MIA is
  defined as the MIA $(P, I, O, \mustderives{}, \mayderives{})$, where
  (i)~$p \mustderives{i} p'$ if $p \derives{i}_P p'$ and $i \in I$,
  and (ii)~$p \mayderives{\alpha} p'$ if $p \derives{\alpha}_P p'$ and
  $\alpha \in I \cup O \cup \{\tau\}$.
\label{def:iaembeddingmia}
\end{defi}

\noindent
In the remainder of this section we simply write~$\embed{p}{}$ for $p
\in \embed{P}{MIA}$.  This embedding is much simpler than the one
of~\cite{LarNymWas2007} since MIA more closely resembles IA than IOMTS
does.  In particular, the following theorem is obvious:

\begin{thm}[IA-Embedding Respects Refinement]
  For IAs~$P, Q$ with $p \in P$, $q \in Q$: $\,p \iasim q$ if and only
  if $\embed{p}{} \miasim \embed{q}{}$.
\label{thm:iaembeddingmia}
\end{thm}


\noindent
Our embedding respects operators~$\andop$ and~$\parop$, unlike the one
in~\cite{LarNymWas2007}:

\begin{thm}[IA-Embedding is a Homomorphism]
  For IAs~$P, Q$ with $p \in P$, $q \in Q$:
  \begin{enumerate}[\bf(a):]
  \item $\embed{p}{} \andop \embed{q}{}$
    $\miaeq$ $\embed{p \andop q}{}$;
  \item $\embed{p}{} \,\parop\, \embed{q}{}$
    $\miaeq$ $\embed{p \parop q}{}$.
  \end{enumerate}
\label{thm:miaembedding}
\end{thm}

\proof{
  Part~(b) follows directly from the definitions of parallel
  composition on IA and MIA, whereas
  Part~(a)''$\sqsupseteq_{\textrm{MIA}}$'' is an immediate consequence
  of Thms.~\ref{thm:miaandisand} and~\ref{thm:iaembeddingmia} by
  general order theory.  We are thus left with proving
  Part~(a)''$\miasim$''.

  Both sides only differ in additional
  transitions~$\mayderives{\alpha}$ with $\alpha \in O \cup \{\tau\}$
  in $\embed{P}{MIA} \andop \embed{Q}{MIA}$, where on the other
  side~$\obsmayderives{\epsilon}\,\mayderives{\alpha}$.  Formally, we
  define the relation $\mc{R} \df \{ \pair{\embed{p}{} \andop
    \embed{q}{}}{\embed{p \andop q}{}} \;|$ $p \in P,\, q \in Q \}
  \cup \text{id}_P \cup \text{id}_Q$ and argue that~$\mc{R}$ is a
  MIA-refinement relation:
  \begin{iteMize}{$\bullet$}
  \item Firstly, $\embed{P}{MIA} \andop \embed{Q}{MIA}$ and $\embed{P
    \andop Q}{MIA}$ are isomorphic on input-transitions since the
    Rules (IMust1)--(IMust3) (and Rules~(IMay1--(IMay3)) exactly
    correspond to Rules (I1)--(I3), as well as on~$P$ and~$Q$.

  \item Secondly, consider a transition $\embed{p}{} \andop
    \embed{q}{} \mayderives{\tau} \embed{p'}{} \andop \embed{q}{}$
    according to Rule~(May1) and $\embed{p}{} \obsmayderivesp{\tau}{P}
    \embed{p'}{}$.  Then, $p \andop q \obsderives{\tau} p' \andop q$
    in IA by repeated application of Rule~(T1) and, therefore,
    $\embed{p \andop q}{} \,\obsmayderives{\tau}\, \embed{p' \andop
      q}{}$ in the IA-embedding.  Rule~(May2) is analogous, and
    Rule~(May3) for $\alpha = \tau$ is similar (with interleaving of
    $\tau$-steps).  In addition, Rule~(May3) for $\alpha \in O$ is
    similar, too, except that the $\tau$-steps are followed by an
    $\alpha$-transition according to Rule~(O).  \qed
  \end{iteMize}
}


\noindent
We observe that the IA-embedding into MIA is `better'
wrt.\ conjunction than that into dMTS since refinement holds in both
directions.  The reason is that MIA-refinement is coarser (i.e.,
larger) than dMTS-refinement applied to MIAs (which are dMTSs after
all): input may-transitions do not have to be matched in the former.
Thus, there can be more lower bounds wrt.\ MIA-refinement and the
greatest lower bound can be larger.

\begin{prop}[Disjunction and IA-Embedding]
  For IAs~$P, Q$ with $p \in P$, $q \in Q$, we have: $\embed{p}{}
  \orop \embed{q}{} \,\miasim\, \embed{p \orop q}{}$.
\label{prop:thm:miaembeddingdisj}
\end{prop}

\noindent
This result holds by general order theory due to
Thm.~\ref{thm:iaembeddingmia}.  The reverse refinement for disjunction
is not valid as we have already seen in
Fig.~\ref{fig:miadisjintuitive}, and this difference repairs a
shortcoming of IA-disjunction as discussed on p.~\pageref{fromtena}.


\section{Conclusions and Future Work}
\label{sec:conclusions}


We introduced \emph{Modal Interface Automata} (MIA), an interface
theory that is more expressive than \emph{Interface Automata}
(IA)~\cite{DeAHen2005}: it allows one to mandate that a
specification's refinement must implement some output, thus excluding
trivial implementations, e.g., one that accepts all inputs but never
emits any output.  This was also the motivation behind
\emph{IOMTS}~\cite{LarNymWas2007} that extends \emph{Modal Transition
  Systems} (MTS)~\cite{Lar89} by inputs and outputs; however, the
IOMTS-parallel operator in the style of IA is not compositional. Apart
from having disjunctive must-transitions, MIA is a subset of IOMTS,
but it has a different refinement relation that is a precongruence for
parallel composition.

Most importantly and in contrast to IA and IOMTS, the MIA theory is
equipped with a conjunction operator for reasoning about components
that satisfy multiple interfaces simultaneously.  Along the way, we
also introduced conjunction on IA and a disjunctive extension of MTS
--~as well as disjunction on IA, MTS and MIA~-- and proved these
operators to be the desired greatest lower bounds (resp., least upper
bounds) and thus compositional.  Compared to the language-based modal
interface theory of~\cite{RacBadBenCaiLegPas2011}, our formalism
supports nondeterministic specifications and allows limited
nondeterminism (in the sense of deterministic \emph{disjunctive}
transitions) even for inputs.  Hence, MIA establishes a theoretically
clean and practical interface theory that fixes the shortcomings of
related work.


\begin{figure}[tbp]
\phantom{.}\hfill%
\includegraphics[scale=0.52, clip=true]{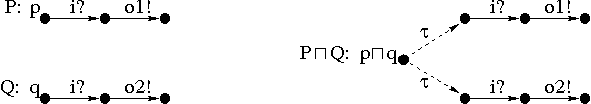}%
\hfill\phantom{.}
\caption{In Logic LTS~\cite{LueVog2010}, disjunction is internal choice.}
\label{fig:intchoice}
\end{figure}

From a technical perspective, our MIA-theory borrows from our earlier
work on Logic LTS~\cite{LueVog2010}.  There, we started from a very
different conjunction operator appropriate for a deadlock-sensitive
CSP-like process theory, and then derived a `best' suitable refinement
relation.  In~\cite{LueVog2010}, disjunction is simply internal
choice~$\sqcap$, as sketched in Fig.~\ref{fig:intchoice}.  For MIA, $p
\sqcap q$ is not suited at all since both~$p$ and~$q$ require that
input~$i$ is performed immediately.


Future work shall follow both theoretical and practical directions.
On the theoretical side, we firstly wish to study MIA's expressiveness
in comparison to other theories via
thoroughness~\cite{FecFruLueSch2009}.  More substantially, however, we
intend to enrich MIA with temporal-logic operators, in the spirit of
truly mixing operational and temporal-logic styles of specification in
the line of our \emph{Logic LTS} in~\cite{LueVog2011}.  Important
guidance for this will be the work of Feuillade and
Pinchinat~\cite{FeuPin2007}, who have introduced a temporal logic for
modal interfaces that is equally expressive to MTS.  In contrast
to~\cite{LueVog2011}, their setting is not mixed, does not consider
nondeterminism, and does not include a refinement relation.  Indeed, a
unique feature of Logic LTS is that its refinement relation subsumes
the standard temporal-logic satisfaction relation.

On the practical side, we plan to study the algorithmic complexity
implied by MIA-refinement, on the basis of existing literature for
MTS.  For example, Antonik et al.~\cite{AntHutLarNymWas2010} discuss
related decision problems such as the existence of a common
implementation; Fisch\-bein and Uchitel~\cite{FisUch2008} generalize
the conjunction of~\cite{LarSteWei95} and study its algorithmic
aspects; Bene\v{s} et al.~\cite{BenCerKre2011} show that refinement
problems for DMTS are not harder than in the case of MTS and also
consider conjunction; Raclet et al.~\cite{RacBadBenCaiLegPas2011}
advocate deterministic automata for modal interface theories in order
to reduce complexity.  In addition, we wish to adapt existing tool
support for interface theories to MIA, e.g., the \emph{MIO
  Workbench}~\cite{BauMaySchHen2010}.


\section*{Acknowledgement}

We thank the anonymous reviewers for their constructive comments and
for pointing out additional related work.  Part of this research was
supported by the DFG (German Research Foundation) under grant
nos.~LU~1748/3-1 and VO~615/12-1 (``Foundations of Heterogeneous
Specifications Using State Machines and Temporal Logic'').


\bibliographystyle{alpha}
\bibliography{literature}

\newcommand{\etalchar}[1]{$^{#1}$}
\begin{thebibliography}{FFELS09}

\bibitem[AHL{\etalchar{+}}10]{AntHutLarNymWas2010}
A.~Antonik, M.~Huth, K.G. Larsen, U.~Nyman, and A.~Wasowski.
\newblock Modal and mixed specifications: {K}ey decision problems and their
  complexities.
\newblock {\em Mathematical Structures in Computer Science}, 20(1):75--103,
  2010.

\bibitem[AL95]{AbaLam95}
M.~Abadi and L.~Lamport.
\newblock Conjoining specifications.
\newblock {\em ACM TOPLAS}, 1(3):507--534, 1995.

\bibitem[BCHS07]{BeyChaHenSes2007}
D.~Beyer, A.~Chakrabarti, T.A. Henzinger, and S.A. Seshia.
\newblock An application of web-service interfaces.
\newblock In {\em {ICWS}}, pages 831--838. IEEE, 2007.

\bibitem[BCK11]{BenCerKre2011}
N.~Bene{\v{s}}, I.~Cern{\'a}, and J.~K{\v{r}}et\'{\i}nsk{\'y}.
\newblock Modal transition systems: {C}omposition and {LTL} model checking.
\newblock In {\em {ATVA}}, volume 6996 of {\em LNCS}, pages 228--242. Springer,
  2011.

\bibitem[BHW11]{BauHenWir2011}
S.~Bauer, R.~Hennicker, and M.~Wirsing.
\newblock Interface theories for concurrency and data.
\newblock {\em Theoret.\ Comp.\ Sc.}, 412(28):3101--3121, 2011.

\bibitem[BJL{\etalchar{+}}12]{BauJuhLarLegSrb2012}
S.~Bauer, L.~Juhl, K.~G. Larsen, A.~Legay, and J.~Srba.
\newblock Extending modal transition systems with structured labels.
\newblock {\em Mathematical Structures in Computer Science}, 22(4):581--617,
  2012.

\bibitem[BMSH10]{BauMaySchHen2010}
S.~Bauer, P.~Mayer, A.~Schroeder, and R.~Hennicker.
\newblock On weak modal compatibility, refinement, and the {MIO} {W}orkbench.
\newblock In {\em {TACAS}}, volume 6015 of {\em LNCS}, pages 175--189.
  Springer, 2010.

\bibitem[CCJK12]{CheChiJonKwi2012}
T.~Chen, C.~Chilton, B.~Jonsson, and M.~Kwiatkowska.
\newblock A compositional specification theory for component behaviours.
\newblock In {\em {ESOP}}, volume 7211 of {\em LNCS}, pages 148--168. Springer,
  2012.

\bibitem[dH01]{DeAHen2001}
L.~{de Alfaro} and T.A. Henzinger.
\newblock Interface automata.
\newblock In {\em {FSE}}, pages 109--120. ACM, 2001.

\bibitem[dH05]{DeAHen2005}
L.~{de Alfaro} and T.A. Henzinger.
\newblock Interface-based design.
\newblock In {\em Engineering Theories of Software-Intensive Systems}, volume
  195 of {\em NATO Science Series}. Springer, 2005.

\bibitem[DHJP08]{DoyHenJobPet2008}
L.~Doyen, T.A. Henzinger, B.~Jobstmann, and T.~Petrov.
\newblock Interface theories with component reuse.
\newblock In {\em {EMSOFT}}, pages 79--88. ACM, 2008.

\bibitem[Dil89]{Dil89}
D.L. Dill.
\newblock {\em Trace Theory for Automatic Hierarchical Verification of
  Speed-Independent Circuits}.
\newblock MIT Press, 1989.

\bibitem[FFELS09]{FecFruLueSch2009}
H.~Fecher, {D. de} Frutos-Escrig, G.~L{\"u}ttgen, and H.~Schmidt.
\newblock On the expressiveness of refinement settings.
\newblock In {\em {FSEN}}, volume 5961 of {\em LNCS}, pages 276--291. Springer,
  2009.

\bibitem[FP07]{FeuPin2007}
G.~Feuillade and S.~Pinchinat.
\newblock Modal specifications for the control theory of discrete event
  systems.
\newblock {\em J.\ Discrete Event Dyn.\ Syst.}, 17:211--232, 2007.

\bibitem[FU08]{FisUch2008}
D.~Fischbein and S.~Uchitel.
\newblock On correct and complete strong merging of partial behaviour models.
\newblock In {\em {SIGSOFT FSE}}, pages 297--307. {ACM}, 2008.

\bibitem[HLL{\etalchar{+}}12]{HatLeaLeinMuePar2012}
J.~Hatcliff, G.~T. Leavens, K.~R.~M. Leino, P.~M{\"u}ller, and M.~Parkinson.
\newblock Behavioral interface specification languages.
\newblock {\em {ACM} Computing Surveys}, 44(3):16, 2012.

\bibitem[Lar90]{Lar89}
K.G. Larsen.
\newblock Modal specifications.
\newblock In {\em Automatic Verification Methods for Finite State Systems},
  volume 407 of {\em LNCS}, pages 232--246. Springer, 1990.

\bibitem[LNW07]{LarNymWas2007}
K.G. Larsen, U.~Nyman, and A.~Wasowski.
\newblock Modal {I/O} automata for interface and product line theories.
\newblock In {\em {ESOP}}, volume 4421 of {\em LNCS}, pages 64--79. Springer,
  2007.

\bibitem[LSW95]{LarSteWei95}
K.G. Larsen, B.~Steffen, and C.~Weise.
\newblock A constraint oriented proof methodology based on modal transition
  systems.
\newblock In {\em {TACAS}}, volume 1019 of {\em LNCS}, pages 17--40. Springer,
  1995.

\bibitem[LV10]{LueVog2010}
G.~L{\"u}ttgen and W.~Vogler.
\newblock Ready simulation for concurrency: {I}t's logical!
\newblock {\em Inform.\ and Comput.}, 208:845--867, 2010.

\bibitem[LV11]{LueVog2011}
G.~L{\"u}ttgen and W.~Vogler.
\newblock Safe reasoning with {Logic LTS}.
\newblock {\em Theoret.\ Comp.\ Sc.}, 412(28):3337--3357, 2011.

\bibitem[LX90]{LarXin90}
K.G. Larsen and L.~Xinxin.
\newblock Equation solving using modal transition systems.
\newblock In {\em {LICS}}, pages 108--117. {IEEE}, 1990.

\bibitem[MB03]{MerBjo2003}
L.~G. Meredith and S.~Bjorg.
\newblock Contracts and types.
\newblock {\em C.\ ACM}, 46(10):41--47, 2003.

\bibitem[Mey92]{Mey92}
B.~Meyer.
\newblock Applying design by contract.
\newblock {\em IEEE Computer}, 25(10):40--51, 1992.

\bibitem[MG05]{MayGru2005}
W.~Maydl and L.~Grunske.
\newblock Behavioral types for embedded software -- {A} survey.
\newblock In {\em Component-Based Software Development}, volume 3778 of {\em
  LNCS}, pages 82--106. Springer, 2005.

\bibitem[RBB{\etalchar{+}}11]{RacBadBenCaiLegPas2011}
J.~Raclet, E.~Badouel, A.~Benveniste, B.~Caillaud, A.~Legay, and R.~Passerone.
\newblock A modal interface theory for component-based design.
\newblock {\em Fund.\ Inform.}, 107:1--32, 2011.

\bibitem[VW02]{VogWol2002}
W.~Vogler and R.~Wollowski.
\newblock Decomposition in asynchronous circuit design.
\newblock In {\em Concurrency and Hardware Design}, volume 2549 of {\em LNCS},
  pages 152--190. Springer, 2002.

\end{thebibliography}


\end{document}